\documentclass[aps,prd,twocolumn,superscriptaddress,nofootinbib]{revtex4-1}

\usepackage{latexsym}
\usepackage{amsmath}
\usepackage{amssymb}
\usepackage{amsfonts}
\usepackage{bm}
\usepackage{physics}
\usepackage{mathrsfs}

\usepackage{color}
\definecolor{purple}{rgb}{0.5,0,0.5}
\definecolor{blue}{rgb}{0.0,0,0.9}
\definecolor{prdblue}{rgb}{0.133,0.118,0.498}
\usepackage[colorlinks=true, pdfstartview=FitV, linkcolor=prdblue, citecolor= prdblue, urlcolor=prdblue]{hyperref}

\usepackage{supertabular} 
\usepackage{placeins}
\usepackage{epsfig}
\usepackage{graphicx}

\begin{document}


\title{Distribution amplitudes of heavy-light pseudo-scalar and vector mesons from Dyson-Schwinger equations framework}


\author{Y.-Z. Xu}
\email[]{yinzhen.xu@dci.uhu.es}
\affiliation{Departmento de Ciencias Integradas, Universidad de Huelva, E-21071 Huelva, Spain.}
\affiliation{Departamento de Sistemas F\'isicos, Qu\'imicos y Naturales, Universidad Pablo de Olavide, E-41013 Sevilla, Spain}


\date{\today}

\begin{abstract}
We report the first Dyson-Schwinger equation predictions for the leading-twist distribution amplitudes of the $B^*$, $B_s^*$, and $B_c^*$ mesons, and extend the investigation to a broader set of heavy-light pseudo-scalar/vector mesons. Numerical analysis shows that, in flavor-asymmetric systems, the distribution amplitude is skewed toward the heavier quark, with the position of the maximum $\sim M^f_E/(M^f_E+M^g_E)$, where $M_E$ is the Euclidean constituent quark mass and $f$, $g$ denote the quark flavors.
 Spin effects are found to be subleading compared to the influence of valence-quark composition, and lead to heavier quarks carrying more light-front momentum in vector mesons than in their pseudo-scalar counterparts. Our results can be compared with both experimental and theoretical outcomes in the future.
\end{abstract}


\maketitle


\section{INTRODUCTION}
\label{sec:intro}

The study of mesons is a fundamental topic in hadron physics. A meson's leading-twist distribution amplitudes (DAs), typically defined as the matrix elements of non-local operators between the meson state and the vacuum, describe the longitudinal momentum distribution of valence quarks in the limit of negligible transverse momentum \cite{Brodsky:1997de,Ball:1998je,Bakulev:2003zf}. Moreover, they are important inputs in many hard exclusive processes at large momentum transfers and play a central role in quantum chromodynamics (QCD) factorization theory \cite{Beneke:1999br,Beneke:2001ev}. In the asymptotic limit, the pion's DAs follow a simple form, $\phi^{\text{asy}}(x)= 6 x(1-x)$ \cite{Lepage:1979zb}. However, their shapes at hadronic scales are non-perturbative and therefore pose a theoretical challenge \cite{LatticeParton:2022zqc}. \par 
The DAs of light mesons, such as the pion and kaon, have been extensively studied in various frameworks \cite{Chang:2013pq,Shi:2014uwa,Serna:2020txe,Lu:2021sgg,Arthur:2010xf,RQCD:2019osh,Zhang:2020gaj}, revealing broadened shapes compared to the asymptotic form, a signal that reflects strong non-perturbative QCD dynamics and relativistic effects. In the case of heavy mesons, although most existing studies focus on the DAs defined in the framework of heavy quark effective theory (HQET) \cite{Kawamura:2001jm}, QCD-based DAs have also attracted growing interest in different approaches, e.g., lattice QCD \cite{Han:2024min}, light-front quark models (LFQM) \cite{Hwang:2010hw, Dhiman:2019ddr}, Algebraic model (AM) \cite{Almeida-Zamora:2023bqb}, and the Dyson-Schwinger/Bethe-Salpeter equations (DSEs/BSEs) framework \cite{Ding:2015rkn,Binosi:2018rht,Serna:2020txe,Serna:2022yfp}. These efforts have significantly advanced our knowledge of the internal dynamics of heavy mesons, extending insights from the light-quark sector to the heavy-quark sector. However, compared with the DAs of heavy pseudo-scalar mesons, studies of their vector counterparts are still relatively scarce.\par  
 Among these approaches, the DSEs/BSEs formalism, which provides a non-perturbative and Poincar\'e-covariant framework capable of simultaneously describing confinement and dynamical chiral symmetry breaking (DCSB), has been successfully applied to the study of hadron properties for over thirty years \cite{Roberts:1994dr,Maris:2003vk,Roberts:2007jh,Qin:2011dd,Rodriguez-Quintero:2010qad,Qin:2011xq,Tandy:2014hja,Xu:2019ilh,Xu:2020loz,Xu:2021lxa,Serna:2021xnr,daSilveira:2022pte,Li:2023zag,Xu:2023izo,Xu:2024fun,Serna:2024vpn,Sultan:2024mva,Xu:2024rew,Raya:2024glv,Miramontes:2024fgo,Chen:2024emt,Xu:2025cyj}. In this framework, the BS wave function is regarded as containing the essential non-perturbative information required to describe the internal structure of a meson. Accordingly, the DAs can be expressed as the light-front projection of the BS wave function and are typically reconstructed from their Mellin moments \cite{Chang:2013pq,Serna:2020txe}. A major numerical challenge in this procedure lies in the evaluation of moments, as the integrands exhibit multi-dimensional oscillatory behavior, particularly for light mesons. Several studies (e.g., Refs.~\cite{Li:2016dzv,Ding:2015rkn}) report that reliable values for them cannot be obtained through a direct approach. Therefore, current calculations of these moments mostly rely on either fitting the Bethe-Salpeter amplitudes (BSAs) using the Nakanishi representation or introducing damping factors for extrapolation \cite{Chang:2013pq,Ding:2015rkn,Lu:2021sgg,Serna:2020txe,Li:2016dzv,Li:2016mah}, although such methods may give rise to certain ambiguities \cite{Serna:2022yfp}. \par 
 The exploration of DAs for heavy mesons within the DSEs/BSEs framework can be traced back to Ref.~\cite{Ding:2015rkn}, which presented a systematic analysis of both pseudoscalar and vector heavy quarkonia. This line of research was advanced in Ref.~\cite{Binosi:2018rht}, where DSE results for the $D$, $D_s$, $B$, $B_s$, and $B_c$ mesons were obtained for the first time. In addition, that work discussed the correspondence between DAs obtained from the BS wave function and those defined in HQET. Further progress was made in Refs.~\cite{Serna:2020txe,Serna:2021xnr}, which provided a novel calculation of these pseudo-scalar heavy-light mesons' DAs using an effective interaction kernel, along with updated and refined results extending the preceding analysis. Most recently, Ref.~\cite{Serna:2022yfp} reported the first DSE predictions for the DAs of the $D^*$ and $D_s^*$ mesons. Their direct computations, which reach up to the 4th-6th Mellin moments depending on the meson, provide the most advanced DSE results available for heavy-light vector mesons. Meanwhile, Ref.~\cite{Shi:2024laj} also extracted the DAs of the $D$, $B$, and $B_c$ mesons from light-front wave functions predicted within the same framework.\par
In this work, we extend these studies by presenting the first DSE predictions for the DAs of the $B^*$, $B_s^*$, and $B_c^*$ mesons, based on our previous investigations on their properties \cite{Xu:2024fun,Xu:2024rew}. For comparison, results for a broader set of heavy-light pseudo-scalar and vector mesons are also provided. To facilitate this analysis, we introduce a novel numerical method, which combines a two-dimensional Chebyshev tensor grid (CTG) \cite{941859,doi:10.1137/130908002} with a multi-dimensional adaptive integration algorithm \cite{Hahn:2004fe,Hahn:2014fua,10.1145/210232.210233,10.1145/210232.210234}, thereby advancing the direct numerical evaluation of moments to the 8th order without extrapolation or fitting. Based on the rich Mellin moment data for these mesons, we reconstruct the corresponding DAs and further investigate their characteristic features, including peak positions and widths, as well as how these properties depend on the valence-quark composition and spin configuration. \par  
This paper is organized as follows: In Sec.\,\ref{sec:2}, we introduce the DSEs/BSEs framework and meson DAs. In Sec.\,\ref{sec:3}, the numerical results of heavy-light pseudo-scalar/vector meson DAs are presented. Based on these results, we discuss the effect of flavor symmetry breaking. Sec.\,\ref{sec:4} provides a brief summary and perspectives.

\section{DSEs/BSEs framework and Meson's Distribution amplitudes}
\label{sec:2}
\subsection{Quark propagators in the complex plane}
We work within the DSEs/BSEs framework in Euclidean space. As a first step, the dressed-quark propagator is obtained from the following gap equation:
\begin{align}
  S^{-1}(k)=\ &Z_2i \gamma \cdot k + Z_4 m + Z_{1} \int^{\Lambda} \frac{d^4 q}{(2 \pi)^4}  g^2 D_{\mu \nu}(k-q) \nonumber \\
   &\frac{\lambda^a}{2} \gamma_\mu S(q) \frac{\lambda^a}{2} \Gamma_\nu(k,q),
  \label{eq.DSE}
\end{align} 
and the general form of $S^{-1}(k)$ is
\begin{equation}
  S^{-1}(k)= i\gamma \cdot pA(k^2) + B(k^2),
\end{equation}
where $A(k^2)$ and $B(k^2)$ are scalar functions, $m$ is current-quark mass, $Z_{1,2,4}$ are the renormalization constants, $\Lambda$ represents a regularization scale. In this work, we employ a mass-independent momentum-subtraction renormalisation scheme \cite{Chang:2008ec,Bashir:2012fs,Liu:2019wzj,Chen:2019otg} and choose renormalization scale $\zeta = 2$ GeV \cite{Xu:2019ilh,Li:2016mah}. The quark mass function $\tilde{M}(k^2)=B(k^2)/A(k^2)$ is independent of $\zeta$ and the renormalisation-group invariant current-quark mass can be defined by \cite{Roberts:2007jh,Chen:2019otg}
\begin{align}
  \hat{m}=\lim _{k^2 \rightarrow \infty}\left[\frac{1}{2} \ln \frac{k^2}{\Lambda_{\mathrm{QCD}}^2}\right]^{\gamma_m} \tilde{M}\left(k^2\right),
  \label{eq.mhat}
\end{align}
accordingly, one-loop evolved current-quark mass reads  \cite{Maris:1997hd,Chen:2019otg}
\begin{align}
   m^{\zeta}=\hat{m}/\left[\frac{1}{2} \ln \frac{\zeta^2}{\Lambda_{\mathrm{QCD}}^2}\right]^{\gamma_m}.
     \label{eq.mloop}
\end{align}
Another quantity that can be obtained from the quark mass function is the Euclidean constituent quark mass \cite{Maris:1997tm}
\begin{align}
  M_E=\left\{k \mid k^2=\tilde{M}^2\left(k^2\right)\right\},
  \label{eq.em}
\end{align}
which provides a realistic estimate of the quark’s active quasi-particle mass \cite{Hecht:2000xa,Xu:2020loz}. \par 
In conjunction with dressed quark-gluon vertex $\Gamma_\nu(k,q) \rightarrow Z_2\gamma_\nu$ \cite{Binosi:2014aea}, which characterizes the rainbow-ladder (RL) truncation, we adopt the following form for the gluon propagator:
\begin{equation}
  Z_1 g^2 D_{\mu \nu}(l)\Gamma_\nu(k,q)=Z_2^2 \mathcal{G}\left(l^2\right) \mathcal{P}_{\mu \nu}^T(l)\gamma_\nu=\mathcal{D}_{\text{eff}}\gamma_\nu\,,
  \label{eq.gluon}
\end{equation}
where $l=k-q$, $\mathcal{P}_{\mu \nu}^T(l)=\delta_{\mu \nu}-{l_\mu l_\nu}/l^2$ is transverse projection operator, and the effective interaction is chosen as the Qin-Chang model\,\cite{Qin:2011dd,Qin:2011xq}
\begin{subequations}
\begin{align}
&\frac{\mathcal{G}(l^2)}{l^2}=\mathcal{G}^{\text{IR}}(l^2)+\frac{8\pi^2\gamma_{m}\mathcal{F}(l^2)}{\ln[\tau+(1+l^2/\Lambda^2_{\text{QCD}})^2]},\\
 &\mathcal{G}^{\text{IR}}(l^2)=D\frac{8\pi^2}{\omega^4}e^{-l^2/\omega^2}\,.
 \end{align}
  \label{eq.qc}
\end{subequations}
with $\mathcal{F}(l^2)=\{1-\exp[(-l^2/(4m_t^2)]\}/l^2$, $m_t=0.5$\,GeV, $\tau=e^2-1$, $\Lambda_{\text{QCD}}=0.234$\,GeV, $\gamma_{m}=12/25$ \cite{Xu:2021mju}. For the infrared model parameters, in line with Refs.\,\cite{Xu:2024fun,Shi:2024laj,Xu:2024rew} we choose $(D\omega)_{u/d}=(0.82\ \text{GeV})^3$, $(D\omega)_s=(0.68\ \text{GeV})^3$, $(D\omega)_c=(0.66\ \text{GeV})^3$ ,$(D\omega)_b=(0.48\ \text{GeV})^3$, with $\omega_{u/d,s}=0.5$ GeV, $\omega_{c,b}=0.8$ GeV. More details of Eqs.\,(\ref{eq.DSE}-\ref{eq.qc}) are presented in Refs. \cite{Roberts:1994dr,Maris:2003vk,Maris:1997tm,Qin:2011dd,Maris:1999nt}.\par 
In the calculation of bound states, arising from the on-shell condition $P^2=-M^2$, the dressed-quark propagator $S(k_\pm)$ is often obtained by solving Eq.\,\eqref{eq.DSE} in the complex plane \cite{Fischer:2005en,Rojas:2014aka}. Here, $M$ denotes the meson mass, $k_\pm = k \pm \alpha_\pm P$, $\alpha_+ = \alpha$, $\alpha_- = 1 - \alpha$, and $\alpha \in [0,1]$ represents the momentum partitioning parameter. In this work, we adopt the common rest frame
\begin{equation}
	k=(0,0,\sin \theta, \cos \theta )|k|,\ P=(0,0,0, iM),
\end{equation}
correspondingly, the value of $k^2_\pm$ is constrained by the following parabola 
\begin{equation}
	\text{Re}(k^2_\pm) = -\alpha^2_\pm M^2 +\frac{\text{Im}^2(k^2_\pm) \sec^2 \theta}{4 M^2 \alpha^2_{\pm}},\   \theta \in [0, \pi].
\end{equation}
Although the physical observables do not depend on $\alpha$, in the actual calculation, the selection of $\alpha$ should be made carefully to avoid the parabola including the pole for the accuracy of the contour integral \cite{Fischer:2005en,Blank:2011qk}. If we define the vertex of parabola as $(- {\Theta}^2, 0)$, the momentum partitioning parameter $\alpha$ should satisfy the following relation \cite{Xu:2024rew}
\begin{align}
  1-\Theta_{\bar{g}}/M<\alpha<\Theta_f/M.
  \label{eq.alpha}
\end{align}
Furthermore, the maximum computable mass is given by $\Theta_f + \Theta_{\bar{g}}$, and the corresponding $\alpha$, namely, the optimal $\alpha = \Theta_f / (\Theta_f + \Theta_{\bar{g}})$ \cite{Rojas:2014aka}. In this work, we obtain $\Theta_u=0.56$ GeV, $\Theta_s=0.67$ GeV, $\Theta_c=1.63$ GeV, $\Theta_b=4.8$ GeV. As long as the quark propagator on the parabola is determined, its value at any interior point of the parabola can be directly obtained through the Cauchy integral theorem \cite{Sanchis-Alepuz:2017jjd}.
\subsection{Meson's BSAs in two-dimensional Chebyshev tensor grid}
\begin{table}
\caption{\label{tab:meson}The masses and decay constants of mesons, with renormalization-group-invariant current-quark mass (see Eq.\,\eqref{eq.mhat}): $\hat{m}_{u/d}=0.0068$ GeV, $\hat{m}_{s}=0.198$ GeV, $\hat{m}_{c}=1.739$ GeV, $\hat{m}_{b}=7.494$ GeV; one-loop evolved current-quark mass in 2 GeV (see Eq.\,\eqref{eq.mloop}): $m^{\zeta_2}_{u/d}=0.0047$ GeV, $m^{\zeta_2}_{s}=0.137$ GeV, $m^{\zeta_2}_{c}=1.205$ GeV, $m^{\zeta_2}_{b}=5.195$ GeV; Euclidean constituent quark mass (see Eq.\,\eqref{eq.em}): $M_E^{u/d}=0.423$ GeV, $M_E^{s}=0.489$ GeV, $M_E^{c}=1.333$ GeV, $M_E^{b}=4.256$ GeV, and $^\dagger$ denote the fitting values from weight factor (see Eq.\,\eqref{eq:kernel.wRL}). For comparison, we collect both experimental values \cite{ParticleDataGroup:2018ovx,CLEO:2000moj} and lQCD's results \cite{Mathur:2018epb,Cichy:2016bci,Dowdall:2012ab,Fu:2016itp,Dudek:2014qha,Donald:2013pea,Lubicz:2017asp,Donald:2012ga,Follana:2007uv,McNeile:2012qf,Bazavov:2017lyh,Colquhoun:2014ica,Colquhoun:2015oha}.}
\begin{ruledtabular}
\begin{tabular}{l|lll|lll}
&\multicolumn{3}{c|}{Mass [GeV]} &\multicolumn{3}{c}{Decay constant [GeV]}\\
  & Expt. & lQCD & Herein &Expt. & lQCD & Herein \\ \hline
$\pi$ & 0.138(1) & - & 0.135  & 0.092(1) & 0.093(1) & 0.095\\
$\rho$ & 0.775(1) & 0.780(16) & 0.755 & 0.153(1)& - & 0.150\\
$\phi$ & 1.019(1) & 1.032(16) & 1.019  & 0.168(1) & 0.170(13) & 0.168\\
$\eta_c$ & 2.984(1) & - & 2.984 & 0.237(52) &0.278(2) &  0.270\\
$J/\psi$ & 3.097(1) & 3.098(3) & 3.114 & 0.294(5) & 0.286(4) & 0.290\\
$\eta_b$ & 9.399(1) & - & 9.399 & - & 0.472(5) & 0.464\\
$\Upsilon$ & 9.460(1) & - & 9.453 & 0.505(4) & 0.459(22) & 0.441\\
  \hline
 $K$ & 0.495(1) &- & 0.495$^\dagger$ &0.110(1)&-& 0.108\\
 $K^*$ & 0.896(1)& 0.993(1) & 0.880 &0.159(1)&-&0.158 \\
 $D$&1.868(1)&1.868(3)& 1.868$^\dagger$  & 0.144(4)&0.150(4)& 0.140  \\
 $D^*$&2.009(1)&2.013(14)& 2.017  &-& 0.158(6)&0.160 \\
 $D_s$&1.968(1)&1.968(4)&1.968$^\dagger$ &0.182(3)&0.177(1)&0.164 \\
 $D^*_s$&2.112(1)&2.116(11)&2.111 &-&0.190(5)&0.186\\
 $B$&5.279(1)&5.283(8)&5.279$^\dagger$ &0.133(18)&0.134(1)&0.123 \\
 $B^*$&5.325(1)&5.321(8)&5.334 &-&0.131(5)&0.126 \\
 $B_s$&5.367(1)&5.366(8)&5.367$^\dagger$ &-&0.163(1)&0.149 \\
 $B^*_s$&5.415(1)&5.412(6)&5.422 &-&0.158(4)&0.151 \\
 $B_c$&6.275(1)&6.276(7)&6.275$^\dagger$&- &0.307(10)&0.300 \\
 $B^*_c$&-&6.331(7)&6.340&-&0.298(9)&0.296 \\
\end{tabular}
\end{ruledtabular}
\end{table}
The next step is the meson's Bethe-Salpeter amplitudes (BSAs). Generically, those can be calculated based on the following homogeneous BSEs
\begin{align}
\Gamma^{f\bar{g}}_H\left(P;k^2, k\cdot P \right)&= \int^{\Lambda} \frac{d^4 q}{(2 \pi)^4}  K^{f\bar{g}}(q,k;P)  S^{f}\left(q_{+}\right)\nonumber \\
& \Gamma^{f\bar{g}}_H\left(P;q^2, q\cdot P \right) S^{g}\left(q_{-}\right),
    \label{eq:hBSE}
\end{align}
with $f$ and $g$ denoting the flavor of (anti-)quark, and the corresponding quark propagators have been discussed in the last subsection. In the RL approximation with a single-flavor gluon model, the interaction kernel is given by \cite{Xu:2021mju}
\begin{equation}
\label{eq:kernel.RL}
  K^{f\bar{g}}(q,k;P) = \tilde{\mathcal{D}}^{f\bar{g}}_{\text{eff}} \frac{\lambda^a}{2} \gamma_\mu \otimes \frac{\lambda^a}{2} \gamma_\nu,\  \tilde{\mathcal{D}}^{f\bar{g}}_{\text{eff}} = \mathcal{D}^f_{\text{eff}},
\end{equation}
where $ \mathcal{D}^f_{\text{eff}}$, defined by Eq.\,\eqref{eq.gluon}, describes the strength of interaction and decreases as the mass of quark increases because the dressed-effect is suppressed \cite{Serna:2017nlr,Qin:2019oar}. \par 
Over the past thirty years, Eq.\,\eqref{eq:kernel.RL} has been widely used in flavor symmetric/slightly asymmetric systems such as $u\bar{d}$, $u\bar{s}$, $c\bar{c}$ \cite{Xu:2019ilh}, guaranteeing that the axial-vector Ward-Green-Takahashi identity (AV-WGTI) is fulfilled. This symmetry preservation ensures pseudo-scalar mesons as Nambu-Goldstone modes and a massless pion in the chiral limit \cite{Munczek:1994zz,Maris:1997hd}. \par 
The situation is however more complicated for heavy-light mesons, where different effective kernels have been applied. In such cases, the AV-WGTI is not automatically preserved for flavor-asymmetric systems. Instead, it can be approximately implemented, and its degree of approximation tested through deviations revealed by the generalized GMOR relation \cite{Chen:2019otg,Serna:2022yfp,Serna:2021xnr,Xu:2024rew}:
\begin{equation}
   f_{0^-} M_{0^{-}} = \left(m_f+m_g\right) \rho_{0^{-}} \,;
   \label{eq.gmor}
\end{equation}
where $M_{0^-}$ corresponds to the pseudo-scalar meson's mass, while $\rho_{0^{-}}$ is defined as follows 
\begin{align}
  \rho_{0^{-}}= Z_4 N_c \text{Tr} \int^{\Lambda}\frac{d^4k}{(2\pi)^4} \gamma_5 S^f\left(k_{+}\right) \Gamma_{0^{-}}^{f\bar{g}}(P;k^2, k\cdot P) S^{\bar{g}}\left(k_{-}\right)\,;
\end{align}
and $f_{0^-}$ denotes the pseudo-scalar meson leptonic decay constant. A more elaborate discussion on this point can be found in Ref. \cite{Maris:1997hd,Rojas:2014aka,Chen:2019otg}.\par 
The treatment of the heavy-light kernel is still an open question \cite{Shi:2024laj}. A general scheme for ensuring that all Ward-Green-Takahashi identities are satisfied has been proposed, e.g., in Refs.\,\cite{Qin:2020jig,Xu:2022kng}; however, as with other ongoing studies exploring flavor dependence, it remains highly challenging. In this work, we adopt the so-called weight-RL approximation, which arithmetically averages the strength of interaction for different flavors as follows: \cite{Qin:2019oar,Xu:2024fun,Shi:2024laj,Xu:2024rew}
\begin{equation}
\label{eq:kernel.wRL}
 \tilde{\mathcal{D}}^{f\bar{g}}_{\text{eff}} = \eta \mathcal{D}^f_{\text{eff}} + (1-\eta)\mathcal{D}^{\bar{g}}_{\text{eff}},
\end{equation}
where a parameter $\eta$ is introduced such that it gives an explicit weight to each flavor contribution. Plainly, in the flavor-symmetric case, the dependence on $\eta$ disappears and Eq.\,\eqref{eq:kernel.RL} is simply recovered. The effect of the weighting parameter has been discussed in Ref. \cite{Qin:2019oar}, and an automatic averaging is therein presented. Herein, following Refs.\,\cite{Xu:2024fun,Shi:2024laj,Xu:2024rew}, we determine the parameter $\eta$ such that the pseudo-scalar mesons' mass is realistically obtained. Then, once $\eta$ is fixed, vector meson masses and vector and pseudo-scalar meson decay constants are well predicted (see Table\,\ref{tab:meson}). \par 
Furthermore, as discussed above, the quality of the approximation based on the effective kernel in Eq.\,\eqref{eq:kernel.wRL} can be assessed by evaluating the deviations shown in Eq.\,\eqref{eq.gmor}. They amount to the level of $\sim 3\%$ \cite{Xu:2024rew}. Moreover, with a strict heavy-light kernel, once the flavor dependence of the interaction is determined in the gap equation, fitting for heavy-light mesons is not required. Therefore, the weighting parameter in our effective kernel represents an additional drawback and cost of calculating heavy-light mesons within the RL framework, introduced to remedy the effects of flavor asymmetry in BSEs.\par 
The general form of $\Gamma(P;k^2, k\cdot P)$ is given by
\begin{equation}
  \Gamma (P;k^2, k\cdot P )=\sum_{i=1}^N \tau^i(k, P) \mathcal{F}_i (k, P),
\end{equation}
where $\tau^i(k,P)$ is basis and $\mathcal{F}_i(k,P)$ is scalar function. For the pseudo-scalar/vector meson, we choose \cite{Qin:2011xq}
\begin{subequations}
\label{eq:basis}
\begin{align}
\tau_{0^{-}}^1&=i \gamma_5,& &\tau_{0^{-}}^3=\gamma_5 \gamma \cdot k k \cdot P, \\
\tau_{0^{-}}^2&=\gamma_5 \gamma \cdot P,& &\tau_{0^{-}}^4=\gamma_5 \sigma_{\mu \nu} k_\mu P_\nu,
\end{align}
and
\begin{align}
& \tau_{1^{-}}^1=i \gamma_\mu^T  \\
& \tau_{1^{-}}^2=i\left[3 k_\mu^T \gamma \cdot k^T-\gamma_\mu^T k^T \cdot k^T\right] \\
& \tau_{1^{-}}^3=i k_\mu^T k \cdot P \gamma \cdot P \\
& \tau_{1^{-}}^4=i\left[\gamma_\mu^T \gamma \cdot P \gamma \cdot k^T+k_\mu^T \gamma \cdot P\right]\\
&\tau_{1^{-}}^5=k_\mu^T, \\
&\tau_{1^{-}}^6=k \cdot P\left[\gamma_\mu^T \gamma^T \cdot k-\gamma \cdot k^T \gamma_\mu^T\right], \\  
&\tau_{1^{-}}^7=\left(k^T\right)^2\left(\gamma_\mu^T \gamma \cdot P-\gamma \cdot P \gamma_\mu^T\right)-2 k_\mu^T \gamma \cdot k^T \gamma \cdot P, \\
&\tau_{1^{-}}^8=k_\mu^T \gamma \cdot k^T \gamma \cdot P.
\end{align}
\end{subequations}
with $V_\mu^T=V_\mu-P_\mu(V \cdot P) / P^2$. Then the decay constant can be obtained easily after normalization of the meson' BSAs \cite{Qin:2011xq}.\par 
With these in hand, the traditional treatment of Eq.\,\eqref{eq:hBSE} is to discretize the integral based on quadrature rule, such as Gauss-Legendre. It then reduces to the eigenvalue problem of the matrix, which can be solved by matrix-free iterative methods, for example, Arnoldi iteration implemented by ARPACK library \cite{Blank:2010bp}. However, the obtained eigenvector $\Gamma_H\left(P;k^2, k\cdot P \right)$ is thus discretized on the $(k^2,z_k)$ plane, where $z_k =  k\cdot P /(|k|\cdot|P|)$. This prompts us to look for a reasonable way to make them continuous. \par  
\begin{figure}
\centering 
\includegraphics[width=.22\textwidth]{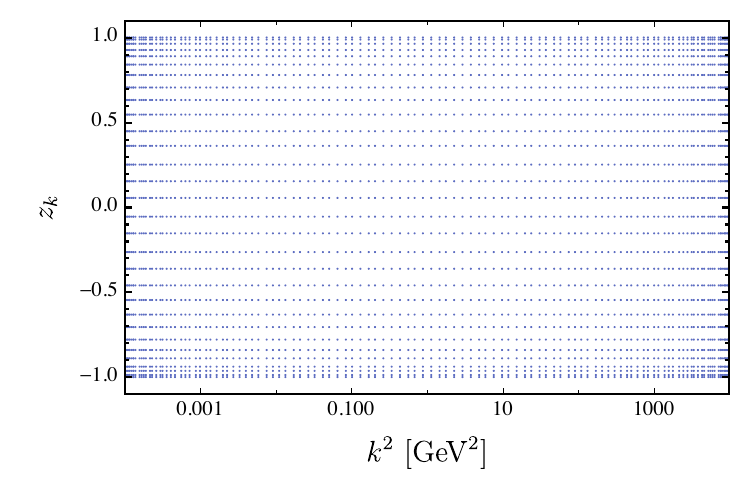}
\includegraphics[width=.22\textwidth]{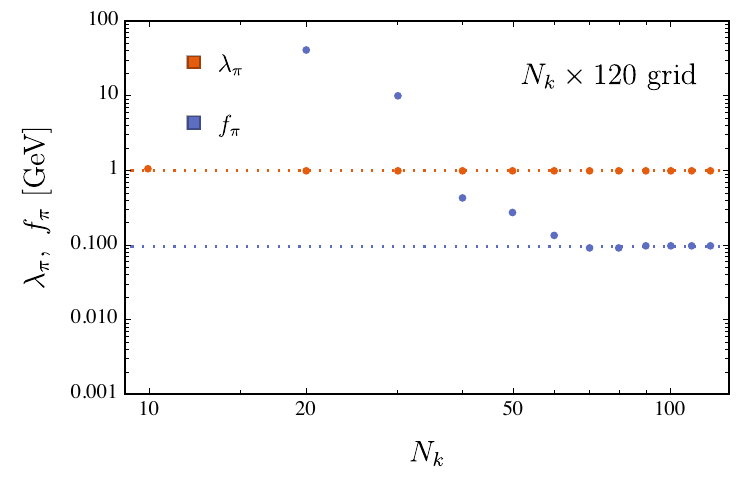}
\includegraphics[width=.22\textwidth]{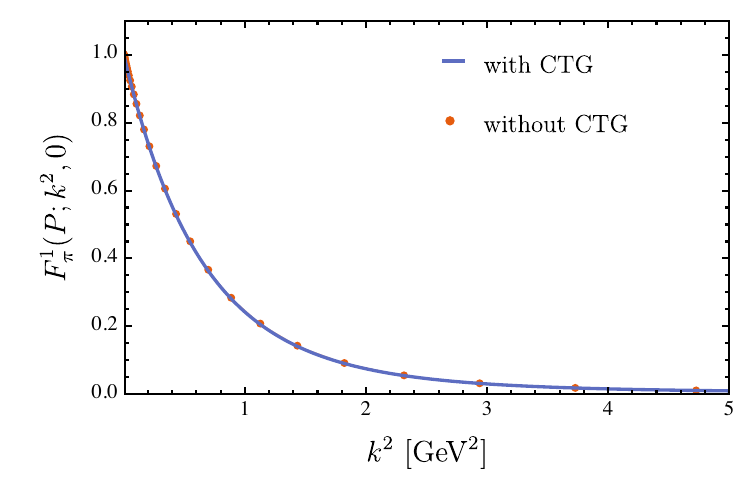}
\includegraphics[width=.22\textwidth]{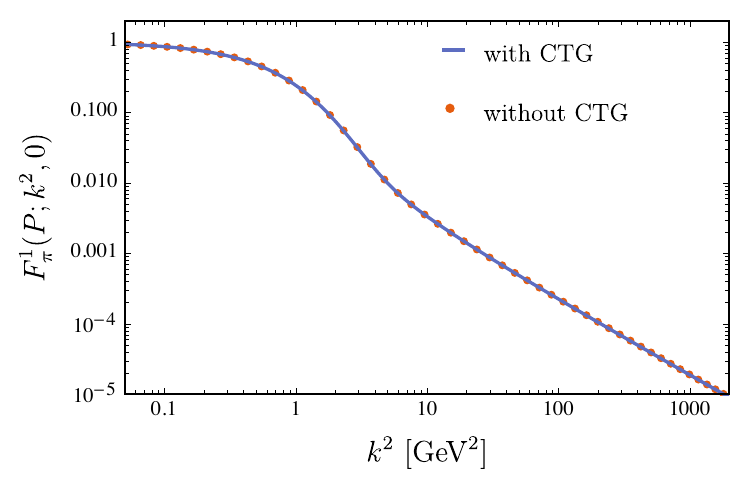}
\caption{\label{fig:chey}The upper left panel illustrates an example of a $N_k \times N_{z_k}$ CTG, with $N_{z_k} = 30$ angular nodes and $N_k = 120$ exponentially mapped radial nodes. The upper right panel shows that as the number of CTG nodes increases, the pion’s eigenvalue $\lambda$ and decay constant $f$ gradually stabilize, approaching the values indicated by the dotted lines, which correspond to those obtained using the traditional method. The lower panels further compare the dominant, unnormalized BSA of the pion calculated by the two approaches.}
\end{figure}
In approximation theory, the Chebyshev series converges for any function that is Lipschitz continuous on $[-1,1]$, and the interpolation polynomial based on Chebyshev nodes provides an approximation that is close to the best polynomial approximation to a continuous function under the maximum norm. In the multi-dimensional case, the tensor products of Chebyshev nodes, namely, the Chebyshev tensor grid (CTG), have been widely used for function reconstruction \cite{941859,doi:10.1137/130908002,glau2020low,PhysRevA.89.013622}. Consider a continuous smooth bivariate function $f: [-1,1]^2 \rightarrow \mathbb{R}$, which can be expanded as
\begin{equation}
f(x, y)=\sum_{i=0}^{\infty} \sum_{j=0}^{\infty} c_{i j} T_i(x) T_j(y),
  \label{eq.chey}
\end{equation}
where $T_n(x)$ is the Chebyshev polynomial of the first kind. According to the discrete orthogonality condition
\begin{equation}
  \sum_{k=0}^{N-1} T_i\left(x_k\right) T_j\left(x_k\right) = \frac{N}{2-\delta_{0i}}\delta_{ij},
\end{equation}
this series can be truncated as
\begin{subequations}
\begin{align}
&f(x, y)\simeq \sum_{i=0}^{N_x-1} \sum_{j=0}^{N_y-1} \tilde{c}_{i j} T_i(x) T_j(y),\\
&\tilde{c}_{ij}=\frac{2-\delta_{0i}}{N_x}\frac{2-\delta_{0j}}{N_y} \sum_{k=0}^{N_x-1} \sum_{k^\prime=0}^{N_y-1} f\left(x_k, y_{k^\prime} \right) T_i\left(x_k\right)T_j\left(y_{k^\prime}\right),
\end{align}
\end{subequations}
where $(x_k, y_{k^\prime})$ are the Chebyshev nodes in the $N_x \times N_y$ grid
\begin{subequations}
\begin{align}
  &(x_k, y_{k^\prime}) = (\cos \left(\frac{2 k+1}{2 N_x} \pi\right),\cos \left(\frac{2 {k^\prime}+1}{2 N_y} \pi\right)), \\
  &k=0,...,N_x-1; {k^\prime} = 0,...,N_y-1.
\end{align}
\end{subequations}
Therefore, the continuous two-dimensional (2D) function can be reconstructed once $f(x_k, y_{k^\prime})$, the function values on the grid of nodes, have been obtained.\par 
Based on this, we map the original $[-1,1]^2$ grid to the actual range we need (see Fig.\,\ref{fig:chey}, upper left panel), and then iterate Eq.\,\eqref{eq:hBSE} on this grid by updating the BSAs at the Chebyshev nodes each time. As the number of nodes increases, the results gradually converge with those obtained using the traditional method (see Fig.\,\ref{fig:chey}, upper right panel). Furthermore, in Fig.\,\ref{fig:chey}, lower panels, we compare the BSAs of the pion with and without the CTG. The results show that both are consistent in the infrared and ultraviolet regions. In this work, we apply a $240 \times 120$ CTG, and the resulting continuous BSAs will be used to calculate the DAs in the next subsection.
\subsection{Distribution amplitudes and Mellin moments}
In the DSEs/BSEs framework, meson DAs $\phi(x)$ can be expressed as the light-front projection of the BS wave function. For the pseudo-scalar meson (PS) \cite{Chang:2013pq,Serna:2020txe,Binosi:2018rht} and the longitudinally/transversely polarized vector meson (VC)\cite{Lu:2021sgg,Serna:2022yfp}, 
\begin{subequations}
\label{eq.DAsdefine}
	\begin{align}
\phi_{0^-}(x)=\ &\frac{N_cZ_2 }{f_{0^-}} \operatorname{Tr}_D \int^{\Lambda} \frac{d^4 k}{(2 \pi)^4} \delta\left(n \cdot k_+-x n \cdot P\right) \nonumber \\
& \gamma_5 \gamma \cdot n \chi\left(k, P\right),\\
\phi_{1^-}^{\|}(x)=\ &\frac{N_c Z_2M_{1^-} }{n \cdot Pf_{1^-}} \operatorname{Tr}_D \int^{\Lambda} \frac{d^4 k}{(2 \pi)^4} \delta\left(n \cdot k_+-x n \cdot P\right) \nonumber\\
&  \gamma \cdot n n_\nu \chi_{\nu}(k, P), \\
\phi_{1^-}^{\perp}(x)=\ &-\frac{N_c Z_T}{2f_{1^-}^{\perp} }  \operatorname{Tr}_D \int^{\Lambda} \frac{d^4 k}{(2 \pi)^4} \delta\left(n \cdot k_+-x n \cdot P\right) \nonumber \\
& n_\mu \sigma_{\mu \rho} \mathcal{O}_{\rho \nu}^{\perp} \chi_{\nu}(k, P),
\end{align}
\end{subequations}
where $M$ is meson's mass with on-shell condition $M^2 = - P^2$, tensor $\mathcal{O}_{\rho \nu}^{\perp}=\delta_{\rho \nu}+n_\rho \bar{n}_\nu+\bar{n}_\rho n_\nu$, $\chi_{H}(k, P)$ denotes the BS wave function, derived from the dressed-quark propagators and BSAs outlined in the preceding subsections:
\begin{equation}
	\chi_{H}(k, P) = S^{f}\left(k_{+}\right) \Gamma_H\left(P;k^2, k\cdot P \right) S^{g}\left(k_{-}\right),
\end{equation}
with renormalization scale $ \mu = 2$ GeV. In this work, we use the Euclidean metric, $ n=(0,0,1, i)$, $\bar{n}=(0,0,-1/2, i/2)$ is a light-like vector and its conjugate. Finally, the normalization conditions $\left\langle x^0\right\rangle = 1$ are constrained by decay constants $f$, $f^{\perp}$ and renormalization constants $Z_{2,T}$. Further discussions on the definition of $\phi(x)$ within the DSEs/BSEs framework, including more details on Eq.\,\eqref{eq.DAsdefine} and its connection with HQET, can be found in Refs. \cite{Chang:2013pq,Tandy:2014hja,Binosi:2018rht,Lu:2021sgg,Serna:2020txe,Serna:2022yfp}.\par 
Due to the presence of the $\delta$ function in Eq.\,\eqref{eq.DAsdefine}, the distribution amplitude $\phi(x)$ is typically reconstructed via its Mellin moments, which are defined as $\langle x^m \rangle = \int_0^1 dx\, x^m \phi(x)$, with explicit expressions given by \cite{Chang:2013pq,Binosi:2018rht,Lu:2021sgg,Serna:2020txe,Serna:2022yfp}
\begin{subequations}
\label{eq.moment}
\begin{align}
\left\langle x^m\right\rangle=\ &\frac{N_cZ_2}{f_{0^-}} \operatorname{Tr}_D \int^{\Lambda} \frac{d^4 k}{(2 \pi)^4} \frac{\left(n \cdot k_+ \right)^m}{(n \cdot P)^{m+1}} \nonumber \\
&\gamma_5 \gamma \cdot n \chi \left(k, P\right),\\
  \left\langle x^m\right\rangle_{\|}=\ &\frac{N_c Z_2M_{1^-} }{f_{1^-}} \operatorname{Tr}_D \int^{\Lambda} \frac{d^4 k}{(2 \pi)^4} \frac{\left(n \cdot k_+\right)^m}{(n \cdot P)^{m+2}} \nonumber \\
 & \gamma \cdot n n_\nu \chi_{\nu}(k, P),\\
\left\langle x^m\right\rangle_{\perp}=\ &-\frac{N_c Z_T}{2 f_{1^-}^{\perp}} \operatorname{Tr}_D \int^{\Lambda} \frac{d^4 k}{(2 \pi)^4} \frac{\left(n \cdot k_+\right)^m}{(n \cdot P)^{m+1}} \nonumber \\
& n_\mu \sigma_{\mu \rho} \mathcal{O}_{\rho \nu}^{\perp} \chi_{\nu}(k, P).
\end{align}
\end{subequations} 
\begin{figure}
\centering 
\includegraphics[width=.22\textwidth]{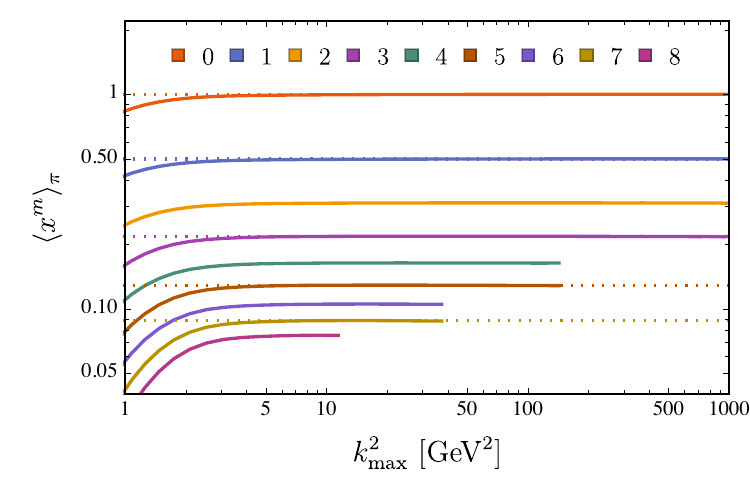}
\includegraphics[width=.22\textwidth]{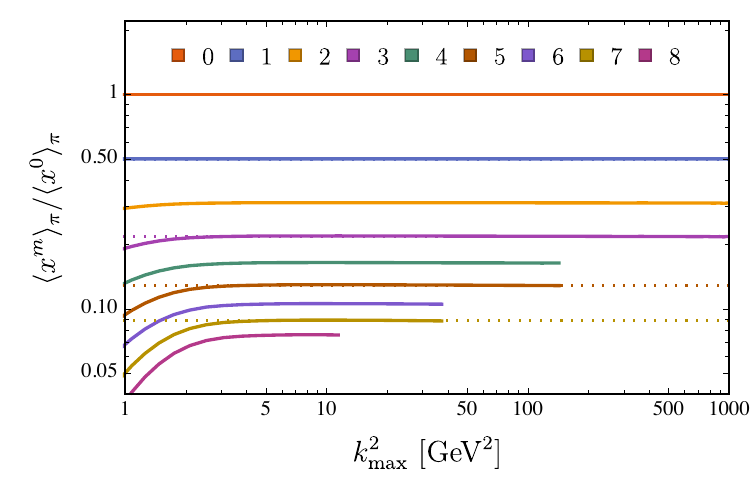}
\includegraphics[width=.22\textwidth]{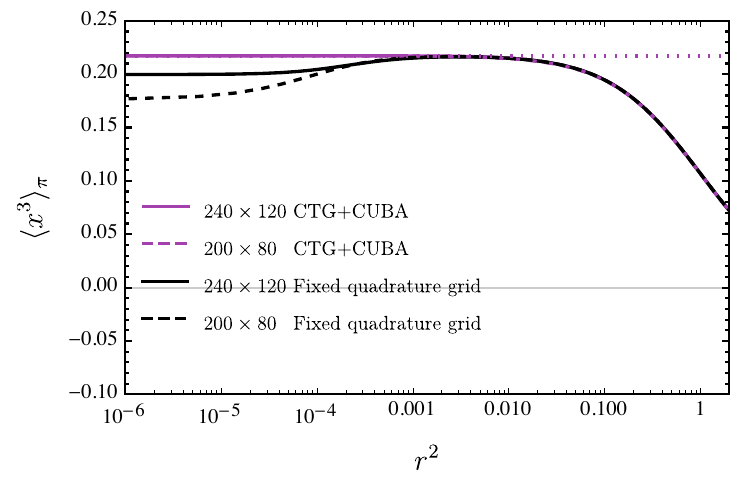}
\includegraphics[width=.22\textwidth]{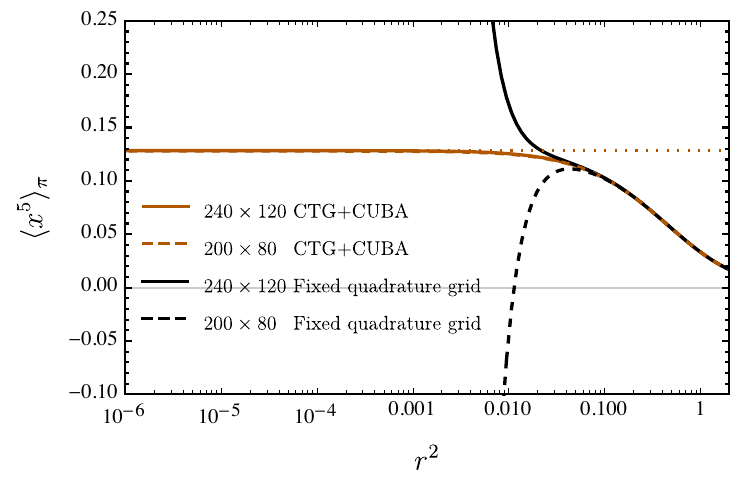}
\caption{\label{fig:check}The upper panels illustrate the variation of $\langle x^m \rangle_\pi$ and $\langle x^m \rangle_\pi / \langle x^0 \rangle_\pi$ with the ultraviolet momentum cutoff $k^2_{\text{max}}$ in Eq.\,\eqref{eq.moment}. Data for higher moments at large cutoffs are unavailable due to numerical instability. In the lower panels, $\langle x^3 \rangle_\pi$ and $\langle x^5 \rangle_\pi$ are used as examples to demonstrate the effect of the damping factor $1 / (1 + k^2 r^2)^m$ in two different $N_k \times N_{z_k}$ CTG. For comparison, results obtained using the traditional fixed quadrature grid are also shown. In all panels, the dotted lines indicate the recurrence results (see Eq.\,\eqref{eq.odd}).}
\end{figure}\par 
In principle, once the BS wave function is known, Mellin moments of arbitrary order can be evaluated. However, as noted in the introduction, the term $\left(n \cdot k_{+}\right)^m$ induces highly multi-dimensional oscillations, which makes the numerical computation of Eq.\,\eqref{eq.moment} challenging, especially for light mesons \cite{Li:2016dzv, Ding:2015rkn}. The rapid development of modern parallel multi-dimensional adaptive integration technology such as CUBA offers new opportunities \cite{Hahn:2004fe,Hahn:2014fua}, which has been successfully applied to similar oscillatory integrals in many fields \cite{Bahr:2018gwf,Longton:2015maa,Kapustin:2010mh}. In the previous subsections, we have obtained continuous quark propagators/BSAs with the help of the Cauchy integral theorem and two-dimensional CTG. Therefore, the CUBA-Cuhre algorithm, a deterministic method grounded in globally adaptive subdivision \cite{10.1145/210232.210233,10.1145/210232.210234}, lends itself naturally to the direct evaluation of Eq.\,\eqref{eq.moment}.\par 
Moreover, the combination of continuous BS wave functions and the CUBA facilitates flexible adjustment of the upper bound of the momentum integration, $k^2_{\rm{max}}$, in Eq.\,\eqref{eq.moment}, a feature that is difficult to realize within the fixed quadrature grid approach. As illustrated in the upper left panel of Fig.\,\ref{fig:check}, using the pion as an example, we examine the behavior of $\langle x^m \rangle$ as a function of the momentum cutoff. It is observed that with increasing $k^2_{\text{max}}$, the integral rapidly rises before gradually approaching a plateau. Furthermore, in practical computations, the ratio $\langle x^m \rangle/\langle x^0 \rangle$ is found to converge to the plateau region more efficiently (see Fig.\,\ref{fig:check}, upper right panel).\par 
For flavor-symmetric ground-state pseudo-scalar and vector mesons, the corresponding DAs satisfy the following symmetry constraint \cite{Li:2016mah}:
\begin{equation}
\int_0^1 dx\, x^m \phi(x) = \int_0^1 dx\, (1-x)^m \phi(x),
\end{equation}
which implies that any odd Mellin moment can be expressed in terms of lower even-order moments, e.g.,
\begin{subequations}
\begin{align}
\langle x^1 \rangle &= \frac{1}{2}, \\
\langle x^3 \rangle &= -\frac{1}{4} + \frac{3}{2} \langle x^2 \rangle, \\
\langle x^5 \rangle &= \frac{1}{2} - \frac{5}{2} \langle x^2 \rangle + \frac{5}{2} \langle x^4 \rangle, \\
\langle x^7 \rangle &= -\frac{17}{8} + \frac{21}{2} \langle x^2 \rangle - \frac{35}{4} \langle x^4 \rangle + \frac{7}{2} \langle x^6 \rangle, \cdots.
\end{align}
\label{eq.odd}
\end{subequations}
This relation provides a stringent consistency check for numerical results. In Fig.\,\ref{fig:check}, the dotted lines indicate moments computed using Eq.\,\eqref{eq.odd}. Once the integrals plateau, our results accurately reproduce these recursion relations.\par 
To further assess the reliability of the calculations, we compare them with results obtained using the extrapolation method. In this approach, a damping factor $1/\left(1 + k^2 r^2\right)^m$ is introduced to suppress oscillations, and the results are extrapolated to $r^2 \to 0$ using data at finite $r^2$ \cite{Li:2016dzv,Li:2016mah}. As illustrated in Fig.\,\ref{fig:check} (lower panel) for $\langle x^3 \rangle$ and $\langle x^5 \rangle$ of the pion, the fixed quadrature grid becomes increasingly unreliable as $r^2$ decreases, exhibiting a noticeable tendency toward divergence and rendering extrapolation unavoidable. In contrast, CTG + CUBA still remains stable.

\section{Numerical results and discussion}
\label{sec:3}
\subsection{Mellin moments}
\begin{table}
\caption{\label{tab:uuPS}First two moments $\langle \xi^{1,2} \rangle$ of $\pi$ and $K$ mesons compared with lattice and DSE results; $\langle \xi^m \rangle$ defined in Eq.\,\eqref{eq.xi}.}
\begin{ruledtabular}
\begin{tabular}{l|l|ll}
    & $\langle \xi^2 \rangle_\pi $  & $\langle \xi \rangle_K $ & $\langle \xi^2 \rangle_K $ \\ \hline
this work  & 0.246 & 0.018 & 0.222\\
lattice(20) \cite{Zhang:2020gaj}  & 0.244(30) & 0.009(18) & 0.198(16)\\
lattice(19) \cite{RQCD:2019osh}  & 0.234(6) & 0.032(17) & 0.231(6)\\
lattice(11) \cite{Arthur:2010xf}  & 0.28(2) & 0.036(2) & 0.26(2)\\
DSE(21) \cite{Lu:2021sgg}  & 0.260 & - & - \\
DSE(20) \cite{Serna:2020txe}  & 0.272(32) & 0.124(13) & 0.234(6)\\
DSE(14,RL,DB) \cite{Shi:2014uwa}  & - & 0.11,0.04 & 0.24,0.23\\
DSE(13,RL,DB) \cite{Chang:2013pq}  & 0.28,0.25 & - & -\\
\end{tabular}
\end{ruledtabular}
\end{table}
With all the above in hand, the Mellin moments of pseudo-scalar/vector mesons can be computed directly. As mentioned before, these moments are defined by
\begin{align}
   \langle x^m \rangle=\int_0^1 d x x^m \phi(x),
\end{align}
and it can be readily transformed to
\begin{align}
  \langle \xi^m\rangle = \langle (2x-1)^m \rangle=\int_0^1 d x (2x-1)^m \phi(x),
  \label{eq.xi}
\end{align}
where $x$ is the momentum fraction carried by the heavier quark. Another commonly discussed moment is the Gegenbauer moment $a_n$ \cite{Ball:2007zt,Braun:2016wnx,Hua:2020gnw}, which arises from the expansion of DAs in terms of Gegenbauer polynomials
\begin{align}
  \phi(x)=6 x(1-x)\left[1+\sum_{n=1}^{\infty} a_n C_n^{3 / 2}(2 x-1)\right].
\end{align}
According to the orthogonality condition
\begin{align}
  &\int^{1}_{0} dx C_n^{3 / 2}(2 x-1)C_m^{3 / 2}(2 x-1) 4x(1-x) \nonumber \\
  &=\frac{(1+n)(2+n)}{(2n+3)}\delta_{nm},
\end{align}
these two types of moments are related by a simple algebraic relation
\begin{align}
  a_n=\frac{2(2 n+3)}{3(n+1)(n+2)} \int_0^1 d x C_n^{3 / 2}(2 x-1) \phi(x).
\end{align}
In particular, for the first and second moment,
\begin{subequations}
\begin{align}
a_1 &= \frac{5}{3}  \langle \xi \rangle = -\frac{5}{3}+\frac{10}{3} \langle x \rangle,\\ 
a_2 &=  -\frac{7}{12}+\frac{35}{12}\langle \xi^2 \rangle = \frac{7}{3}-\frac{35}{3} \langle x \rangle+\frac{35}{3}\langle x^2 \rangle.
\end{align}
\label{eq.gegen}
\end{subequations}
\begin{table*}
\caption{\label{tab:uuVC}First two moments $\langle \xi^{1,2} \rangle$ of $\rho$ and $K^*$ mesons compared with lattice, sum rule, and DSE results; $\langle \xi^m \rangle$ defined in Eq.\,\eqref{eq.xi}. The symbol $ ^\prime $ denotes results derived from the Gegenbauer moments reported in the main text or supplementary materials of their paper, following Eq.\,\eqref{eq.gegen}.}
\begin{ruledtabular}
\begin{tabular}{l|ll|llll}
    & $\langle \xi^2 \rangle^{\|}_\rho $  & $\langle \xi^2 \rangle^{\perp}_\rho $ & $\langle \xi \rangle^{\|}_{K^*} $ & $\langle \xi^2 \rangle^{\|}_{K^*} $  & $\langle \xi \rangle^{\perp}_{K^*} $ & $\langle \xi^2 \rangle^{\perp}_{K^*} $\\ \hline
this work  & 0.259 & 0.236 & 0.023 & 0.220 & 0.033 & 0.210 \\
lattice(21)$^\prime$ \cite{Hua:2020gnw}  & - & - & 0.003(4) & 0.205(3) & 0.044(4) & 0.262(4)\\
lattice(17)$^\prime$ \cite{Braun:2016wnx}   & 0.245(9) & 0.235(8) & - & - & - & -\\
lattice(10) \cite{Arthur:2010xf}  & 0.27(2) & - & 0.043(3) & 0.25(2) & - & -\\
DSE(22) \cite{Serna:2022yfp}  & 0.263 & 0.250 & 0.018 & 0.272 & 0.056 & 0.298\\
DSE(21) \cite{Lu:2021sgg}  & 0.224(12) & 0.240(4) & - & - & - & -\\
DSE(14) \cite{Gao:2014bca}  & 0.23 & 0.25 & - & - & - & -\\
sum rule$^\prime$ \cite{Ball:2007zt}  & 0.234(17) & 0.238(17) & 0.0012(12) & 0.227(21) & 0.0018(18) & 0.227(21) \\
\end{tabular}
\end{ruledtabular}
\end{table*}
\begin{table*}[!t]
\caption{\label{tab:hlm}First moments $\langle \xi \rangle$ of flavor-asymmetric mesons compared with other approaches’ results; $\langle \xi^m \rangle$ is defined in Eq.\,\eqref{eq.xi}.}
\begin{ruledtabular}
\begin{tabular}{l|lllllll}
\multicolumn{1}{l|}{}   & \multicolumn{1}{c}{this work} & \multicolumn{1}{c}{DSE(19) \cite{Binosi:2018rht}} & \multicolumn{1}{c}{DSE(20) \cite{Serna:2020txe}}  & \multicolumn{1}{c}{DSE(22) \cite{Serna:2022yfp}} & \multicolumn{1}{c}{LFQM(10) \cite{Hwang:2010hw}} & \multicolumn{1}{c}{LFQM(19) \cite{Dhiman:2019ddr}} & \multicolumn{1}{c}{AM \cite{Almeida-Zamora:2023bqb}} \\ \hline
  $K$& 0.018 & - & 0.124(13) & - & - & - & -\\
 $K^{*\|}$& 0.023 & - & - & 0.018 & - & - & -\\
  $K^{*\perp}$& 0.033 & - & - & 0.056 & - & - & - \\
  $D$& 0.288 & 0.36(2) & 0.266(17) & - &  0.288,0.251 &0.325 &0.365(13)\\
 $D^{*\|}$& 0.331 & - & - & 0.388 & - &0.356  & -\\
  $D^{*\perp}$& 0.356 & - & - & 0.484 & - & 0.351 & - \\
   $D_s$  & 0.271 & 0.34(2)  &  0.156(18) & - & 0.213,0.207 & 0.311  & 0.335(14) \\
  $D_s^{*\|}$& 0.308 & - & - & 0.254 & - & 0.323 & -\\
  $D_s^{*\perp}$& 0.330 & - & - & 0.310 & - & 0.321 & -\\
  $B$& 0.608 & 0.62(2) & 0.666(10) & -&  0.617,0.531 &0.665 &  0.616(20)\\
   $B^{*\|}$& 0.656 & - & - & - &-  & 0.672  & -\\
  $B^{*\perp}$& 0.671 & - & - & - & - & 0.672  & -\\
  $B_s$& 0.594 & 0.60(2) & 0.642(6) & -  & 0.549,0.486 & 0.651 &0.589(24)\\
    $B_s^{*\|}$& 0.629 & - & - & - & -&0.652  & -\\
  $B_s^{*\perp}$& 0.653 & - & - & - & -&0.653  & -\\
  $B_c$ & 0.374& 0.42(2) & 0.464(4) & - & 0.536,0.368& - &0.413(35) \\
    $B_c^{*\|}$& 0.405 & - & - & - & - & -  & -\\
  $B_c^{*\perp}$& 0.416 & - & - & - & - & - & -\\
\end{tabular}
\end{ruledtabular}
\end{table*}
We first calculate the Mellin moments of the pion and kaon, which have been extensively studied using various methods. A comparison with lattice QCD and previous DSE results is presented in Table~\ref{tab:uuPS}. For $\langle \xi^2 \rangle_{\pi,K}$, different approaches yield consistent values. However, the results of $\langle \xi \rangle_K$ reported by DSE (RL) are significantly larger, because the interaction kernel strength, fitted to pion properties alone is not optimal in the treatment of heavier quark \cite{Xu:2019ilh}. Therefore, it is overestimated for kaon and leading to an excessive $u-s$ quark splitting \cite{Shi:2014uwa}, which is also observed in the gravitational form factors of kaon \cite{Xu:2023izo}. In contrast, the weight-RL scheme suppresses the $s$-quark contribution, yielding predictions closer to lattice QCD and DSE (DB) results.\par 
Similarly, for the $\rho$ meson, our results align well with previous studies, while small deviations appear in the $K^*$ case (see Table~\ref{tab:uuVC}). Specifically, our second moment for the $K^*$ meson is lower than that of DSE (22) \cite{Serna:2022yfp}, but agrees better with QCD sum rule predictions \cite{Ball:2007zt}. Besides, most studies find minor differences between longitudinal and transverse moments \cite{Ball:2007zt,Serna:2022yfp}, except for Ref.~\cite{Hua:2020gnw}, which reports a notable splitting. Given the limited studies on $K^*$ mesons, further studies from different approaches are warranted.\par
\begin{figure}
\centering 
\includegraphics[width=.43\textwidth]{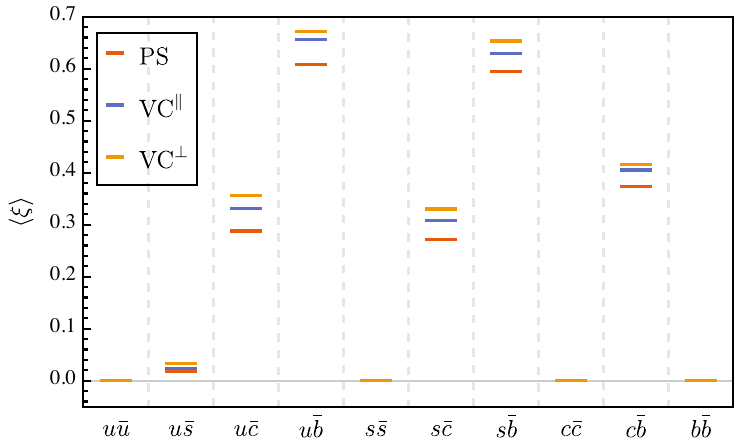}
\caption{\label{fig:xi1}The first moments $ \langle \xi \rangle $ of vector mesons and their pseudo-scalar partner. A universal pattern, i.e, Eq.\,\eqref{eq.pattern}, can be clearly observed.}
\end{figure}

 
Heavy-light mesons such as $u\bar{c}$, $u\bar{b}$, and $c\bar{s}$ exhibit stronger flavor asymmetry than light mesons, offering deeper insight into QCD dynamics. This asymmetry can be reflected in their first moments. In Table~\ref{tab:hlm}, we compare these moments of flavor-asymmetric mesons obtained in this work with previous DSE results, alongside predictions from other models. Numerical results indicate that higher flavor asymmetry corresponds to a larger $\langle \xi \rangle$. Furthermore, it is found that the effect of valence quark spin-flip is smaller than that of flavor asymmetry. For systems with identical valence quark content, we observe a universal pattern (see Fig.\,\ref{fig:xi1}): 
\begin{align}
  \langle \xi \rangle_{0^-} < \langle \xi \rangle^{\|}_{1^-} < \langle \xi \rangle^{\perp}_{1^-},
  \label{eq.pattern}
\end{align}
suggesting heavier quarks carrying more light-front momentum in vector mesons than in their pseudo-scalar counterparts. We note that similar results have also been reported in the LFQM (19)~\cite{Dhiman:2019ddr}; however, the first moments of the longitudinal and transverse components of vector mesons in their work do not follow the same pattern found in our analysis.\par 
The full set of the first eight moments is listed in Appendix~\ref{appendix}. In the next subsection, we will reconstruct their DAs using these data and provide further discussion.
\subsection{Distribution amplitudes}
Due to numerical difficulties, in DSEs/BSEs framework, current DAs reconstructions mainly rely on fitting Mellin moments using prior models. For heavy mesons, a widely used ansatz is \cite{Serna:2020txe,Serna:2022yfp,Ding:2015rkn,Binosi:2018rht,Almeida-Zamora:2023bqb}
\begin{equation}
\phi(x; \alpha,\beta) = \mathcal{N}(\alpha,\beta) 4x(1-x) e^{4\alpha x(1-x)+\beta (2x-1)},
\label{eq.fit1}
\end{equation}
where $\alpha$ and $\beta$ are fitting parameters, and $\mathcal{N}(\alpha,\beta)$ ensures normalization, $\int_0^1 dx\, \phi(x) = 1$. Although $\mathcal{N}(\alpha,\beta)$ has an analytical expression involving special functions, its value should be more reliably obtained via numerical integration to avoid floating-point errors.\footnote{For example, with $\alpha = 0.2$ and $\beta = 5$, the analytical result yields $\mathcal{N} = 0.2179$ in double precision and $\mathcal{N} = 0.1895$ in quadruple precision, whereas numerical integration consistently gives $\mathcal{N} = 0.1895$ in both cases.}\par 
To maintain consistency, Eq.~\eqref{eq.fit1} is uniformly adopted in this work. Its reliability is assessed by using 3 out of 8 moments as input, generating $C_8^3 = 56$ subsets to construct the error band. Additionally, we apply another commonly used model \cite{Chang:2013pq, Gao:2014bca} to the $\pi$ and $\rho$ mesons:
\begin{align}
 \phi(x;a) = \mathcal{N}(a) x^a (1-x)^a,
\end{align}
denoted as fit2 for comparison. Both models yield satisfactory fits, although fit1 exhibits slight fluctuations around $x \sim 0.5$.\par
\begin{figure}
\centering 
\includegraphics[width=.22\textwidth]{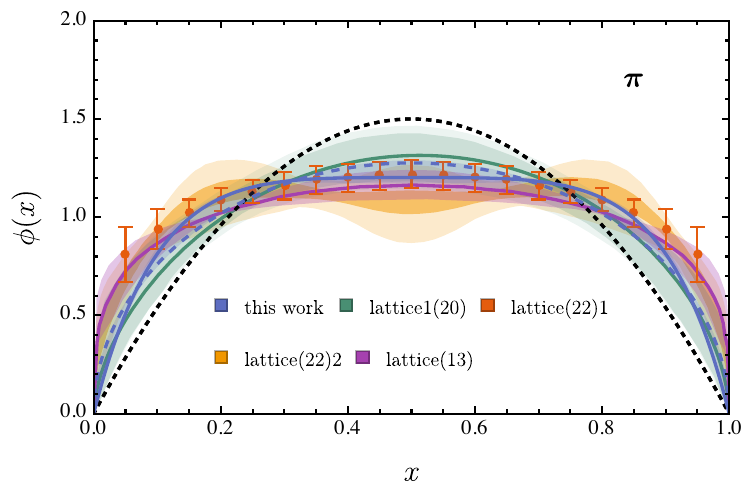}
\includegraphics[width=.22\textwidth]{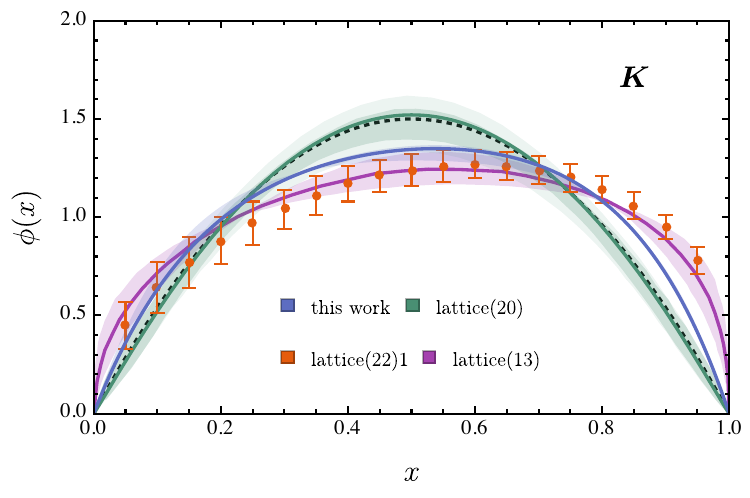}
\caption{\label{fig:fit_test}
The DAs of $\pi$ and $K$ mesons compared with recent lattice results: lattice(13) \cite{Segovia:2013eca}, lattice(20) \cite{Zhang:2020gaj}, lattice(22)1 \cite{LatticeParton:2022zqc} and lattice(22)2 \cite{Gao:2022vyh}. Solid lines represent fit1, dashed lines correspond to fit2, the black dotted lines mark the asymptotic limit. We remind that the first DSE predictions for these two mesons have already been provided in Refs.~\cite{Shi:2014uwa,Chang:2013pq}.}
\end{figure}

\begin{figure}
\centering 
\includegraphics[width=.22\textwidth]{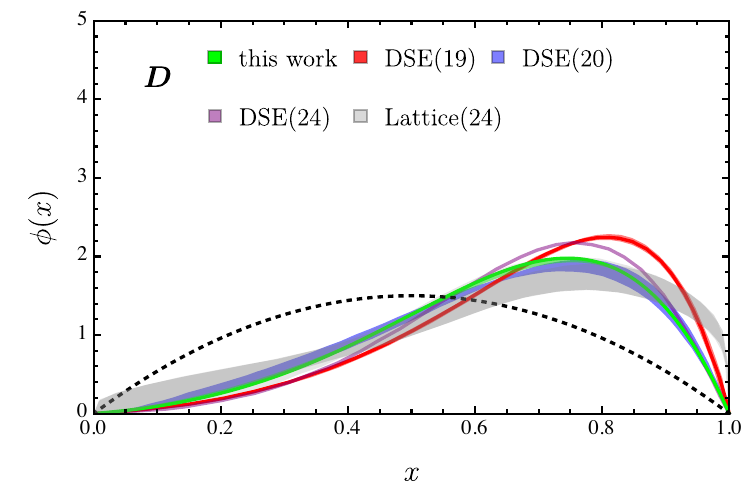}
\includegraphics[width=.22\textwidth]{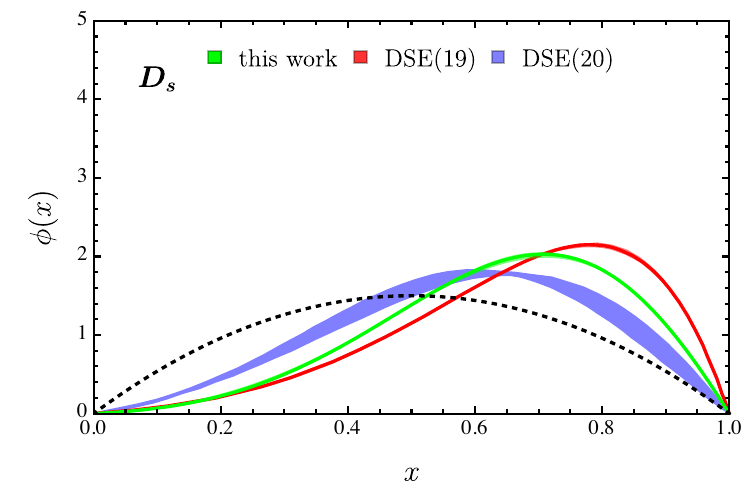}
\includegraphics[width=.22\textwidth]{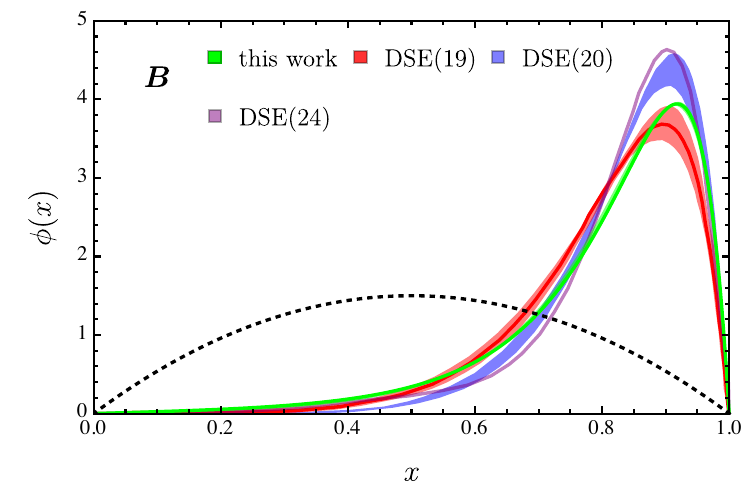}
\includegraphics[width=.22\textwidth]{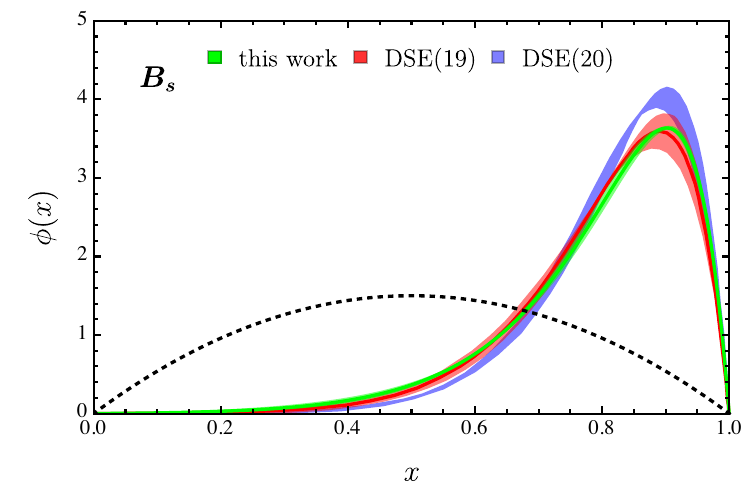}
\caption{\label{fig:hlcom}
The DAs of some heavy-light pseudo-scalar mesons compared with recent lattice and DSE results: DSE(19) \cite{Binosi:2018rht}, DSE(20) \cite{Serna:2020txe}, DSE(24) \cite{Shi:2024laj} and lattice(24) \cite{Han:2024min}. The black dotted lines mark the asymptotic limit.}
\end{figure}

\begin{figure*}
\centering 
\includegraphics[width=.32\textwidth]{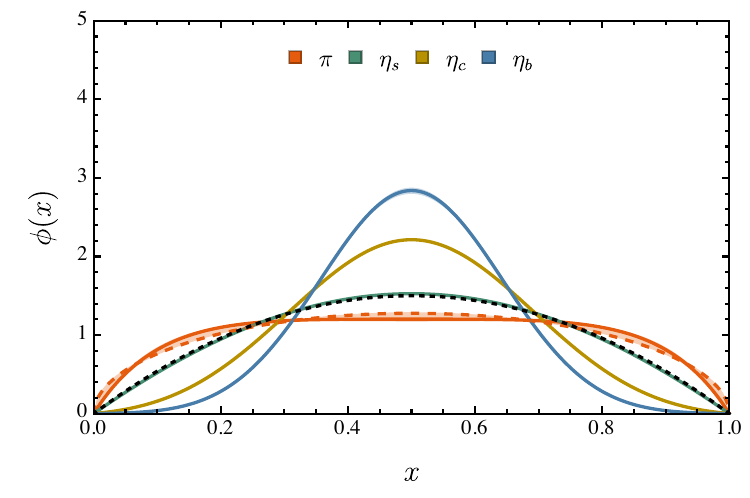}
\includegraphics[width=.32\textwidth]{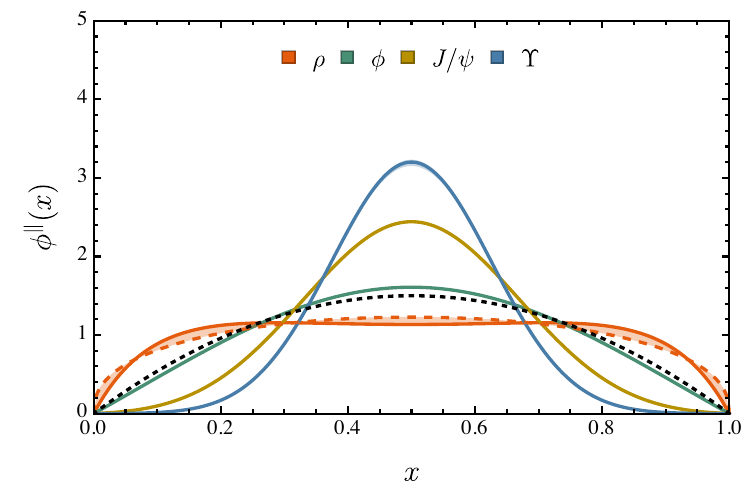}
\includegraphics[width=.32\textwidth]{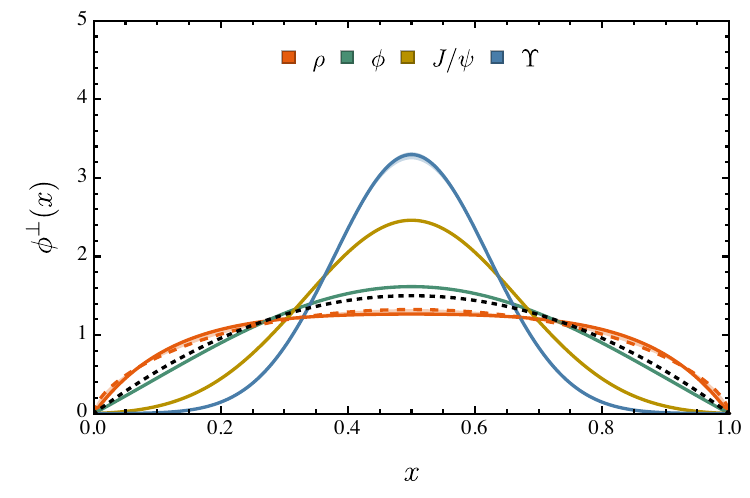}
\includegraphics[width=.32\textwidth]{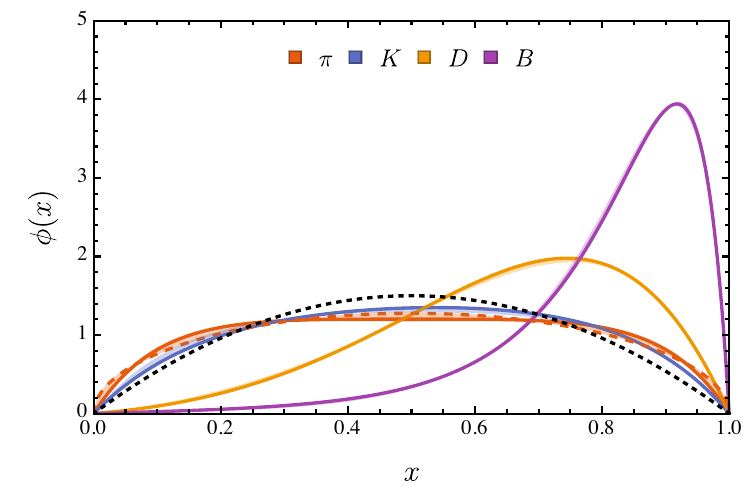}
\includegraphics[width=.32\textwidth]{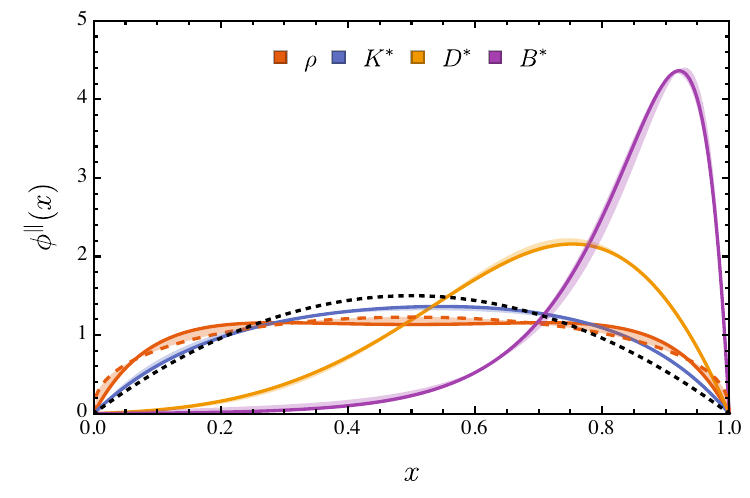}
\includegraphics[width=.32\textwidth]{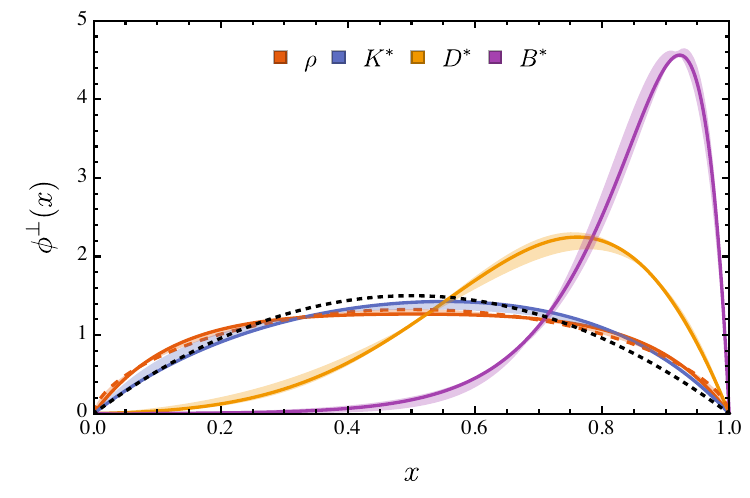}
\includegraphics[width=.32\textwidth]{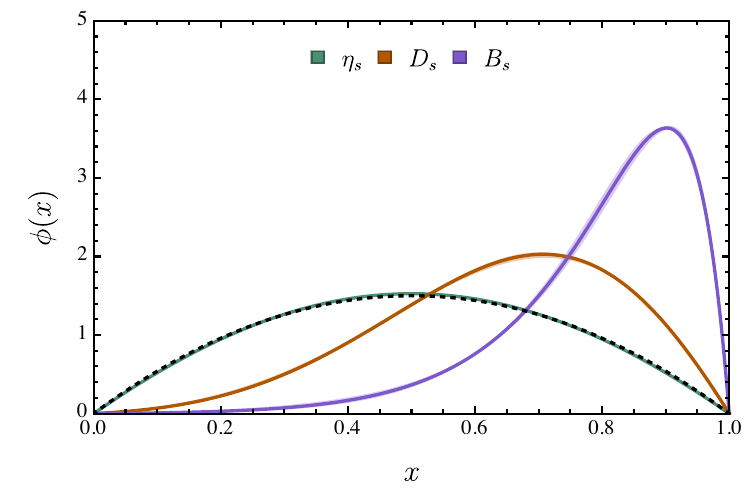}
\includegraphics[width=.32\textwidth]{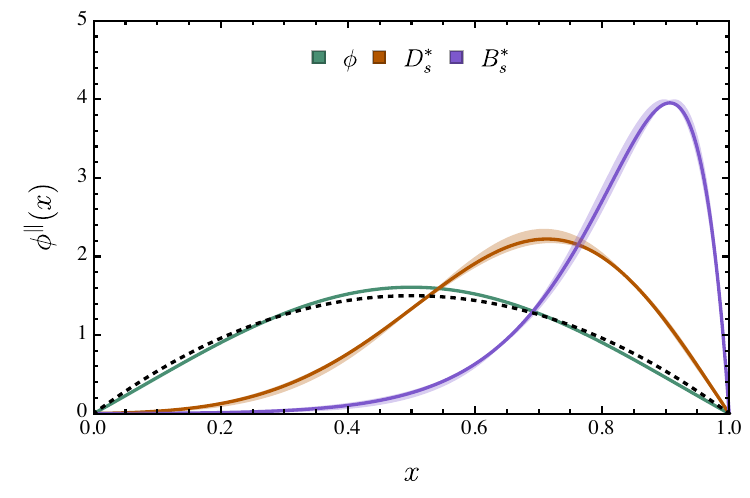}
\includegraphics[width=.32\textwidth]{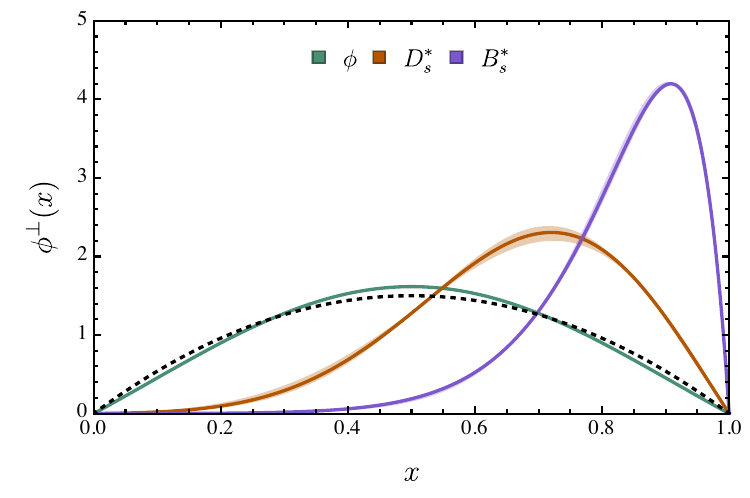}
\includegraphics[width=.32\textwidth]{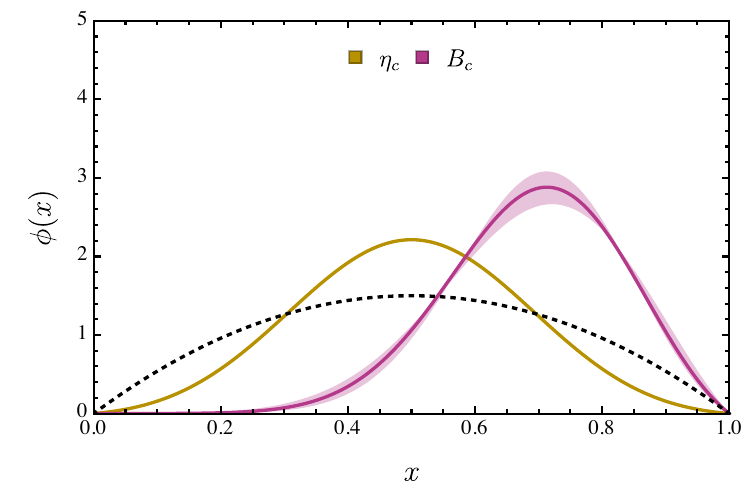}
\includegraphics[width=.32\textwidth]{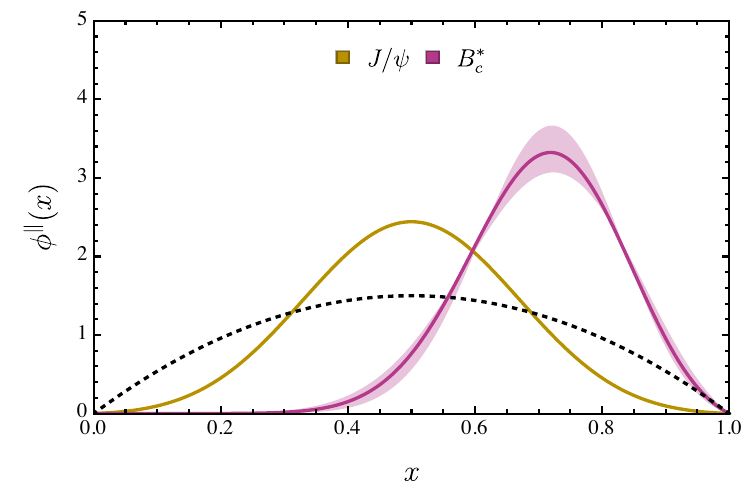}
\includegraphics[width=.32\textwidth]{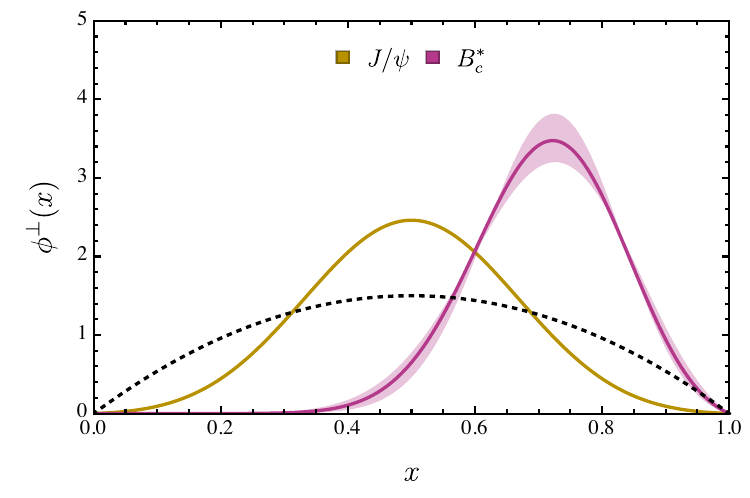}
\caption{\label{fig:full} The DAs of vector mesons and their pseudo-scalar partners. Solid lines represent fit1, dashed lines correspond to fit2, the black dotted lines mark the asymptotic limit.}
\end{figure*}

 
In Figs.~\ref{fig:fit_test} and \ref{fig:hlcom}, we present our results for some pseudo-scalar mesons and compare them with available lattice data and previous DSE studies. These results exhibit qualitative consistency, which is encouraging, particularly for heavy-light mesons, where different DSE studies adopt varying kernel treatments. Since studies on the DAs of vector mesons are relatively scarce, corresponding comparisons are not provided here. The complete set of our DAs results is shown in Fig.~\ref{fig:full}.\par 
At tree level, and in leading order of the expansion in the relative velocities, the quark and the antiquark in the non-relativistic (NR) wave function simply share the momentum of the meson according to their masses \cite{Bell:2008er}, and DAs may have the following form:
\begin{align}
  \phi^{\text{NR}}(x) \simeq \delta\left(x-x^\prime\right),\ x^\prime=m_1/(m_1+m_2),
  \label{eq.nr}
\end{align}
where $x$ is the light-cone momentum fraction of the quark, and $m_{1,2}$ are the (anti-)quark masses. It is thus reasonable to expect that, for ground-state pseudo-scalar and vector flavor-symmetric $q\bar{q}$ mesons, $\phi(x)$ broadens around $x = 1/2$ as the current-quark mass decreases due to enhanced dressing effects. This hypothesis has been consistently supported by previous DSE studies \cite{Ding:2015rkn, Serna:2020txe, Serna:2022yfp}, despite variations in widths. Our results further corroborate this picture (see Fig.~\ref{fig:full}, upper panels). Compared with the asymptotic form $\phi^{\text{asy}}(x)$, the DAs of the $\pi$ and $\rho$ mesons are broader, while those of the $\phi$ meson are slightly narrower, with $\phi_\phi^{\|}(x, \mu) \approx \phi_\phi^{\perp}(x, \mu)$. Notably, a similar conclusion for the $\phi$ meson was also reported in DSE(22), although it remains sensitive to the current-quark mass \cite{Serna:2022yfp}.\par
\begin{table*}[!t]
\caption{\label{tab:FWHM}This table presents an analysis of meson DAs (see Fig.\,\ref{fig:full}), and we work in the isospin-symmetric limit $u = d$. Here, $x_0$ denotes the position of the maximum, $\Delta_x$ represents the FWHM, and $x_{M_E}$ is defined in Eq.\,\eqref{eq.M_E}. The reconstruction of DAs predominantly employs fit1, except for $\pi$ and $\rho$, which utilize fit2 due to fluctuations around $x \sim 0.5$ observed under fit1. If $\Delta_x$ is redefined as the full width at $\left.\phi(x)/2\right|_{x=0.5}$ for fit1, the results are $\Delta^{\pi}_{x}=0.875^{+0.013}_{-0.013},\ \Delta^{\rho^{\|}}_{x}=0.902^{+0.007}_{-0.007},\ \Delta^{\rho^{\perp}}_{x}=0.843^{+0.009}_{-0.009}$, which are approximately consistent with those obtained using fit2. An intuitive comparison is illustrated in Fig.\,\ref{fig:FWHM}.
}
\begin{ruledtabular}
\begin{tabular}{l|l|l|lll|lll}
\multicolumn{2}{c|}{Meson} &\multicolumn{1}{c|}{$x_{M_E}$} &\multicolumn{3}{c|}{$x_0$} & \multicolumn{3}{c}{$\Delta_x$} \\
\multicolumn{2}{c|}{} & \multicolumn{1}{c|}{}   & \multicolumn{1}{c}{PS} & \multicolumn{1}{c}{$\text{VC}^{\|}$} & \multicolumn{1}{c|}{$\text{VC}^{\perp}$} &   \multicolumn{1}{c}{PS} & \multicolumn{1}{c}{$\text{VC}^{\|}$} & \multicolumn{1}{c}{$\text{VC}^{\perp}$} \\ \hline
   $\pi,\rho$ & $u\bar{u}$& 0.500 & $0.500^{+0.000}_{-0.000}$ &  $0.500^{+0.000}_{-0.000}$ &  $0.500^{+0.000}_{-0.000}$ & $0.863^{+0.010}_{-0.011}$ & $0.903^{+0.003}_{-0.004}$ &  $0.826^{+0.005}_{-0.005}$ \\
  $K,K^*$ & $u\bar{s}$& 0.536 & $0.538^{+0.004}_{-0.029}$&  $0.546^{+0.002}_{-0.016}$ &  $0.552^{+0.002}_{-0.033}$ & $0.798^{+0.034}_{-0.010}$ &  $0.790^{+0.028}_{-0.006}$ &  $0.751^{+0.061}_{-0.009}$\\
   $D,D^*$ & $u\bar{c}$& 0.759 &  $0.745^{+0.008}_{-0.003}$ &  $0.753^{+0.004}_{-0.006}$ &  $0.761^{+0.016}_{-0.007}$ &   $0.500^{+0.014}_{-0.003}$ &  $0.455^{+0.004}_{-0.015}$ & $0.436^{+0.031}_{-0.011}$ \\
  $B,B^*$ &$u\bar{b}$& 0.910 &  $0.918^{+0.002}_{-0.005}$ & $0.921^{+0.009}_{-0.004}$ & $0.922^{+0.007}_{-0.014}$ &   $0.216^{+0.007}_{-0.004}$ & $0.201^{+0.006}_{-0.015}$ &  $0.195^{+0.012}_{-0.010}$ \\
   $\eta_s,\phi$ &   $s\bar{s}$ & $0.500$&   $0.500^{+0.000}_{-0.000}$ &  $0.500^{+0.000}_{-0.000}$ &  $0.500^{+0.000}_{-0.000}$  & $0.692^{+0.002}_{-0.001}$ & $0.647^{+0.005}_{-0.002}$ & $0.643^{+0.003}_{-0.002}$\\
  $D_s,D^*_s$ & $s\bar{c}$ & 0.731  & $0.707^{+0.003}_{-0.003}$ & $0.713^{+0.006}_{-0.004}$ & $0.720^{+0.004}_{-0.004}$   &$0.489^{+0.012}_{-0.007}$   & $0.443^{+0.012}_{-0.026}$   & $0.426^{+0.021}_{-0.016}$ \\
 $B_s,B^*_s$ & $s\bar{b}$ & 0.897 &  $0.902^{+0.005}_{-0.006}$ & $0.907^{+0.007}_{-0.009}$ & $0.908^{+0.004}_{-0.005}$   &$0.246^{+0.005}_{-0.005}$   & $0.229^{+0.006}_{-0.008}$   & $0.218^{+0.003}_{-0.003}$ \\
  $\eta_c,J/\psi$ &$c\bar{c}$ & 0.500&  $0.500^{+0.000}_{-0.000}$ &  $0.500^{+0.000}_{-0.000}$ &  $0.500^{+0.000}_{-0.000}$  &$0.436^{+0.002}_{-0.001}$ & $0.391^{+0.002}_{-0.002}$ &  $0.388^{+0.003}_{-0.001}$ \\
  $B_c,B^*_c$ & $c\bar{b}$ & 0.762 &  $0.713^{+0.008}_{-0.002}$ & $0.719^{+0.004}_{-0.002}$ & $0.722^{+0.004}_{-0.002}$    & $0.335^{+0.030}_{-0.023}$  & $0.288^{+0.026}_{-0.028}$   & $0.275^{+0.026}_{-0.026}$ \\
 $\eta_b,\Upsilon$ &$b\bar{b}$ & 0.500 & $0.500^{+0.000}_{-0.000}$ &  $0.500^{+0.000}_{-0.000}$ &  $0.500^{+0.000}_{-0.000}$  & $0.334^{+0.005}_{-0.004}$ & $0.295^{+0.005}_{-0.003}$ &  $0.286^{+0.005}_{-0.003}$\\
\end{tabular}
\end{ruledtabular}
\end{table*}
\begin{figure*}
\centering 
\includegraphics[width=.43\textwidth]{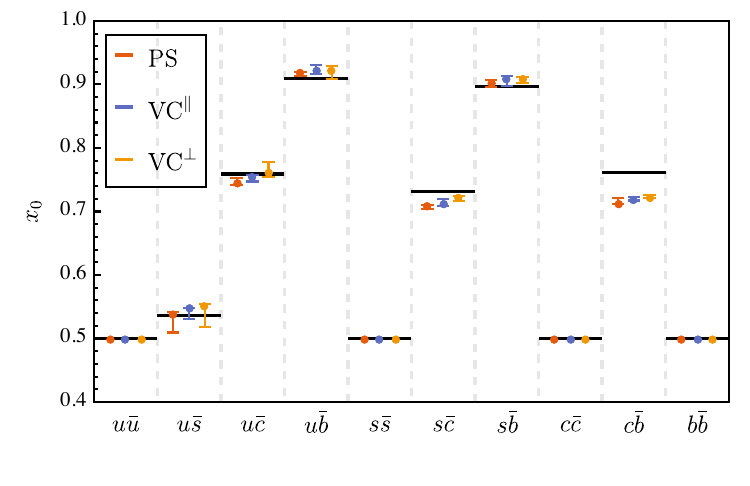}
\includegraphics[width=.43\textwidth]{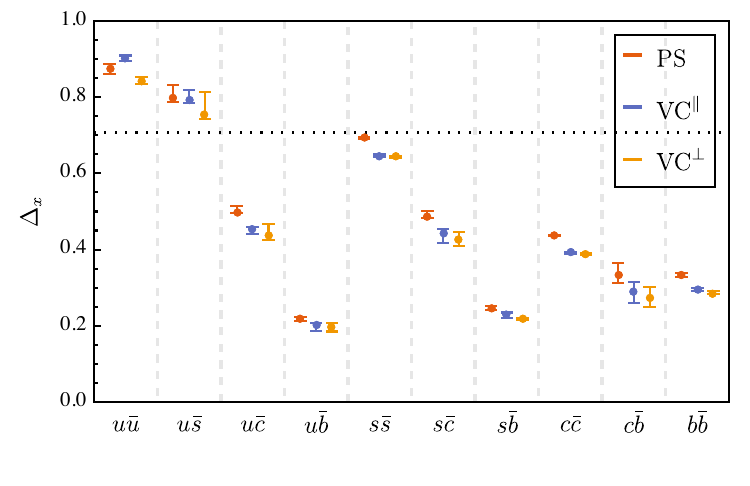}
\caption{\label{fig:FWHM}This figure illustrates an analysis of meson DAs (see Fig.\,\ref{fig:full}). The left panel depicts the position of the maximum $x_0$, with black lines indicating $x_{M_E}$ (see Eq.\,\eqref{eq.M_E}). The right panel shows the full width at half maximum (FWHM) $\Delta_x$, where the black dotted line represents the FWHM of the asymptotic limit ($\sim 0.707$). More details can be found in Table\,\ref{tab:FWHM}.}
\end{figure*}\par 
For heavy-light systems, we observe that their DAs gradually narrow and become increasingly skewed to one side as flavor asymmetry increases (see Fig.\,\ref{fig:full}). To better reflect the variation of DAs with the current-quark mass and flavor asymmetry, we extract two quantities: the position of the maximum, $x_0$, and the full width at half maximum (FWHM), $\Delta_x$ (see Table\,\ref{tab:FWHM} and more intuitive Fig.\,\ref{fig:FWHM}). As noted below Eq.\,\eqref{eq.em}, the Euclidean constituent quark mass $M_E$ serves as a realistic estimate of the quark’s active quasi-particle mass. Inspired by Eq.\,\eqref{eq.nr}, we define
\begin{align}
  x_{M_E} = M^f_E/(M^f_E+M^g_E),
  \label{eq.M_E}
\end{align}
where $f$ and $g$ represent the heavier and lighter quark flavors in the meson, respectively. This definition allows us to compare $x_{M_E}$ with the position of the maximum for DAs, $x_0$. \par 
Interestingly, our results show that $x_0$ and $x_{M_E}$ nearly coincide, except for a deviation of approximately $7\%$ in the $c\bar{b}$ system (see, Fig.\,\ref{fig:FWHM}, left panel). Considering the larger error bands of reconstructed DAs near $x_0$ for $B^{(*)}_c$ meson (see the last row of Fig.\,\ref{fig:full}), this suggests that $\phi(x)$ for ground-state pseudo-scalar and vector mesons may be approximated as a finite-width expansion of the $\delta$-function centered at $x_{M_E}$. The right panel of Fig.\,\ref{fig:FWHM} shows that the width $\Delta_x$ of $u\bar{q}$, $s\bar{q}$, $c\bar{q}$ systems generally decrease with increasing current mass of $q=u, s, c, b$ quark. For all systems except $u\bar{u}$ and $u\bar{s}$, $\Delta_x$ are below the asymptotic limit ($\sim 0.707$).  \par 
When the valence quark composition of flavor-asymmetric systems is identical, we observe $x^{0^-}_0 \lesssim x^{1^-,\|}_0 \lesssim x^{1^-,\perp}_0$ (see Fig.\,\ref{fig:FWHM}, left panel), consistent with the conjecture based on the first moment analysis in the previous subsection. However, the effect of valence quark spin-flip is smaller than that of flavor asymmetry, as reported in the studies of the electromagnetic form factor \cite{Xu:2024fun}. We note that the DAs of $K^*$, $D^*$, and $D^*_s$ mesons have been investigated in Ref. \cite{Serna:2022yfp} with a comparable approach, revealing $x^{1^-,\|}_0 \lesssim x^{1^-,\perp}_0$. Our results support this conclusion, although the effective heavy-light kernel applied is different. Additionally, it is found that $\Delta_x^{0^-} \gtrsim\Delta_x^{1^-,\|}  \gtrsim  \Delta_x^{1^-,\perp}$ for flavor-asymmetric mesons (see, Fig.\,\ref{fig:FWHM}, right panel). However, this relation should be approached with caution due to the larger error bars and further studies are required.
\section{Summary and Perspectives}
\label{sec:4}
In this work, we report the first DSEs/BSEs predictions for the DAs of the $B^*$, $B_s^*$, and $B_c^*$ mesons, and extend the investigation to a broader set of heavy-light vector mesons and their pseudo-scalar partners. To facilitate this analysis, a novel numerical method is introduced. The key step involves iterating the BSE on a two-dimensional Chebyshev tensor grid to obtain a continuous BS wave function, followed by a modern multi-dimensional adaptive integration algorithm to evaluate highly oscillatory integrals. This approach enables the direct computation of Mellin moments up to the 8th order without extrapolation/fitting.\par 
The numerical results show that as the current mass of the $q = u, s, c, b$ quark increases, the DAs of flavor-symmetric $q\bar{q}$ mesons gradually narrow, consistent with previous DSE studies. For flavor-asymmetric $u\bar{q}$, $s\bar{q}$, and $c\bar{q}$ mesons, the DAs not only narrow but also become increasingly skewed, with their maxima located near $M_E^f / (M_E^f + M_E^g)$, where $M_E$ is the Euclidean constituent quark mass and $f/g$ denote the heavier/lighter quark flavors, respectively. Except for the $u\bar{u}$ and $u\bar{s}$ systems, the full width at half maximum (FWHM), $\Delta_x$, remains below the asymptotic limit ($\sim 0.707$).\par 
Furthermore, comparing these DAs reveals that the effect of valence-quark spin-flip is smaller than that of flavor asymmetry. For flavor-asymmetric systems with identical valence-quark composition, the first Mellin moments satisfy $\langle \xi \rangle_{0^-} < \langle \xi \rangle^{\|}_{1^-} < \langle \xi \rangle^{\perp}_{1^-}$, where $\xi = 2x - 1$ and $x$ denotes the momentum fraction carried by the heavier quark. The reconstructed DAs also show maxima positions following $x_0^{0^-} \lesssim x_0^{1^-,\|} \lesssim x_0^{1^-,\perp}$, indicating that heavier quarks carrying more light-front momentum in vector mesons than in their pseudo-scalar counterparts. In addition, the FWHM values satisfy $\Delta_x^{0^-} \gtrsim \Delta_x^{1^-,\|} \gtrsim \Delta_x^{1^-,\perp}$ for flavor-asymmetric mesons; however, this relation should be approached with caution due to the larger error bars and further studies are required.\par 
Our predictions can be compared with future experimental and theoretical results. Moreover, since the proposed numerical method for computing moments imposes no constraints on the integrand, it has broader applications, such as for other meson DAs or the reliable evaluation of Mellin moments of light-front wave functions, which also suffer from similar oscillatory behavior \cite{Serna:2024vpn,Shi:2024laj}.\par 
Looking ahead, as the DAs in this study are reconstructed via an ansatz fit, developing more sophisticated numerical strategies for direct extraction from moment data remains a compelling direction for future investigation. In parallel, we employ an effective interaction within the RL framework; hence, systematic comparisons with alternative kernels and improvements through more advanced beyond-RL schemes are warranted. We expect that the present analysis will contribute to a deeper understanding of the internal structure and emergent dynamics of QCD bound states. 
\appendix
\section{Fitting DAs based on Mellin moments}
\label{appendix}

\begin{table*}[!t]
\caption{\label{tab:PS_all} First eight order Mellin moments $\langle x^m \rangle$ of pseudo-scalar mesons, where $\alpha$, $\beta$ are fit1's parameters, and the fit2's parameter $a_\pi$ is $0.506^{+0.028}_{-0.025}$.}
\begin{ruledtabular}
\begin{tabular}{l|llllllll|rl}
  & $\langle x \rangle $ & $\langle x^2 \rangle $ & $\langle x^3 \rangle $ & $\langle x^4 \rangle $ & $\langle x^5 \rangle $ & $\langle x^6 \rangle $ & $\langle x^7 \rangle $& $\langle x^8 \rangle $ &  \multicolumn{1}{c}{$\alpha$} &  \multicolumn{1}{c}{$\beta$}  \\ \hline
$\pi$ & 0.500 & 0.311 & 0.217 & 0.163 & 0.128 & 0.105 & 0.088 & 0.075 & $-0.986^{+0.094}_{-0.101}$ & $0$\\
$K$ & 0.509 & 0.315 & 0.216 & 0.159 & 0.123 & 0.098 & 0.081 & 0.067 & $-0.510^{+0.058}_{-0.189}$ & $0.076^{+0.006}_{ -0.064}$\\
$\eta_s$ & 0.500 & 0.299 & 0.199 & 0.141 & 0.105 & 0.082 & 0.065 & 0.053 & $0.086^{+0.005}_{-0.011}$ & $0$\\
$D$ & 0.644 & 0.452 & 0.336 & 0.260 & 0.207 & 0.169 & 0.142 & 0.119 & $0.382^{+0.066}_{-0.215}$ & $1.666^{+0.030}_{ -0.136}$\\
$D_s$ & 0.635 & 0.440 & 0.321 & 0.243 & 0.191 & 0.154 & 0.127 & 0.105 & $0.902^{+0.114}_{-0.167}$ & $1.750^{+0.060}_{ -0.106}$\\
$\eta_c$ & 0.500 & 0.279 & 0.169 & 0.109 & 0.074 & 0.052 & 0.038 & 0.028 & $2.530^{+0.026}_{-0.030}$ & $0$\\
$B$ & 0.804 & 0.670 & 0.570 & 0.495 & 0.433 & 0.382 & 0.342 & 0.307 & $-1.419^{+0.476}_{-0.176}$ & $3.171^{+0.478}_{ -0.181}$\\
$B_s$ & 0.797 & 0.657 & 0.553 & 0.473 & 0.410 & 0.358 & 0.318 & 0.284 & $-0.464^{+0.503}_{-0.475}$ & $3.780^{+0.515}_{ -0.498}$\\
$B_c$ & 0.687 & 0.494 & 0.366 & 0.278 & 0.214 & 0.169 & 0.135 & 0.110 & $4.161^{+0.986}_{-1.093}$ & $4.586^{+0.833}_{ -0.771}$\\
$\eta_b$ & 0.500 & 0.269 & 0.153 & 0.092 & 0.058 & 0.038 & 0.025 & 0.018 & $5.167^{+0.173}_{-0.192}$ & $0$\\
\end{tabular}
\end{ruledtabular}
\end{table*}

\begin{table*}[!t]
\caption{\label{tab:PX_all} First eight order Mellin moments $\langle x^m \rangle^{\|} $ of vector meson (longitudinally), where $\alpha$, $\beta$ are fit1's parameters, and the fit2's parameter $a_{\rho^{\|}}$ is $0.409^{+0.009}_{-0.007}$.}
\begin{ruledtabular}
\begin{tabular}{l|llllllll|rl}
  & $\langle x \rangle^{\|} $ & $\langle x^2 \rangle^{\|}  $ & $\langle x^3 \rangle^{\|}  $ & $\langle x^4 \rangle^{\|}  $ & $\langle x^5 \rangle^{\|}  $ & $\langle x^6 \rangle^{\|}  $ & $\langle x^7 \rangle^{\|}  $& $\langle x^8 \rangle^{\|}  $ &  \multicolumn{1}{c}{$\alpha$} &  \multicolumn{1}{c}{$\beta$}  \\ \hline
$\rho$ & 0.500 & 0.315 & 0.223 & 0.169 & 0.134 & 0.110 & 0.093 & 0.080 & $-1.212^{+0.066}_{-0.069}$ & $0$\\
$K^*$ & 0.512 & 0.317 & 0.218 & 0.160 & 0.123 & 0.099 & 0.080 & 0.067 & $-0.470^{+0.030}_{-0.152}$ & $0.100^{+0.006}_{ -0.053}$\\
$\phi$ & 0.500 & 0.296 & 0.194 & 0.136 & 0.100 & 0.077 & 0.062 & 0.049 & $0.363^{+0.015}_{-0.030}$ & $0$\\
$D^*$ & 0.666 & 0.478 & 0.358 & 0.279 & 0.223 & 0.183 & 0.151 & 0.128 & $0.787^{+0.296}_{-0.102}$ & $2.154^{+0.224}_{ -0.064}$\\
$D^*_s$ & 0.654 & 0.460 & 0.338 & 0.258 & 0.202 & 0.162 & 0.132 & 0.109 & $1.474^{+0.513}_{-0.244}$ & $2.297^{+0.375}_{ -0.130}$\\
$J/\psi$ & 0.500 & 0.275 & 0.163 & 0.102 & 0.067 & 0.045 & 0.032 & 0.023 & $3.439^{+0.053}_{-0.037}$ & $0$\\
$B^*$ & 0.828 & 0.702 & 0.605 & 0.527 & 0.465 & 0.413 & 0.372 & 0.338 & $-0.983^{+0.512}_{-1.248}$ & $4.145^{+0.550}_{ -1.339}$\\
$B^*_s$ & 0.814 & 0.680 & 0.579 & 0.502 & 0.437 & 0.385 & 0.341 & 0.305 & $-0.179^{+1.061}_{-0.676}$ & $4.513^{+1.226}_{ -0.673}$\\
$B^*_c$ & 0.703 & 0.510 & 0.380 & 0.288 & 0.222 & 0.174 & 0.138 & 0.110 & $6.312^{+1.924}_{-1.398}$ & $6.610^{+1.775}_{ -1.106}$\\
$\Upsilon$ & 0.500 & 0.265 & 0.147 & 0.086 & 0.052 & 0.033 & 0.021 & 0.014 & $6.930^{+0.161}_{-0.250}$ & $0$\\
\end{tabular}
\end{ruledtabular}
\end{table*}

\begin{table*}[!t]
\caption{\label{tab:CZ_all} First eight order Mellin moments $\langle x^m \rangle^{\perp}$ of vector meson (transversely), where $\alpha$, $\beta$ are fit1's parameters, and the fit2's parameter $a_{\rho^{\perp}}$ is $0.605^{+0.013}_{-0.013}$.}
\begin{ruledtabular}
\begin{tabular}{l|llllllll|rl}
  & $\langle x \rangle^{\perp}  $ & $\langle x^2 \rangle^{\perp} $ & $\langle x^3 \rangle^{\perp} $ & $\langle x^4 \rangle^{\perp} $ & $\langle x^5 \rangle^{\perp} $ & $\langle x^6 \rangle^{\perp} $ & $\langle x^7 \rangle^{\perp} $& $\langle x^8 \rangle^{\perp} $ &  \multicolumn{1}{c}{$\alpha$} &  \multicolumn{1}{c}{$\beta$}  \\ \hline
$\rho$ & 0.500 & 0.309 & 0.214 & 0.158 & 0.123 & 0.099 & 0.082 & 0.070 & $-0.772^{+0.054}_{-0.061}$ & $0$\\
$K^*$ & 0.517 & 0.319 & 0.219 & 0.159 & 0.123 & 0.097 & 0.079 & 0.066 & $-0.269^{+0.052}_{-0.313}$ & $0.153^{+0.006}_{ -0.122}$\\
$\phi$ & 0.500 & 0.296 & 0.194 & 0.136 & 0.100 & 0.077 & 0.061 & 0.049 & $0.389^{+0.013}_{-0.022}$ & $0$\\
$D^*$ & 0.678 & 0.491 & 0.371 & 0.290 & 0.233 & 0.190 & 0.160 & 0.135 & $0.925^{+0.296}_{-0.649}$ & $2.405^{+0.209}_{ -0.492}$\\
$D^*_s$ & 0.665 & 0.471 & 0.349 & 0.267 & 0.209 & 0.168 & 0.138 & 0.115 & $1.680^{+0.352}_{-0.399}$ & $2.571^{+0.248}_{ -0.308}$\\
$J/\psi$ & 0.500 & 0.275 & 0.162 & 0.101 & 0.066 & 0.045 & 0.032 & 0.023 & $3.518^{+0.025}_{-0.063}$ & $0$\\
$B^*$ & 0.835 & 0.715 & 0.619 & 0.545 & 0.484 & 0.429 & 0.384 & 0.346 & $-0.623^{+2.354}_{-0.958}$ & $4.818^{+2.925}_{ -1.028}$\\
$B^*_s$ & 0.827 & 0.698 & 0.598 & 0.519 & 0.456 & 0.401 & 0.358 & 0.321 & $0.318^{+0.735}_{-0.460}$ & $5.413^{+0.869}_{ -0.508}$\\
$B^*_c$ & 0.708 & 0.517 & 0.386 & 0.293 & 0.226 & 0.177 & 0.140 & 0.112 & $7.097^{+2.002}_{-1.563}$ & $7.420^{+1.877}_{ -1.278}$\\
$\Upsilon$ & 0.500 & 0.264 & 0.146 & 0.084 & 0.051 & 0.032 & 0.020 & 0.014 & $7.426^{+0.165}_{-0.314}$ & $0$\\
\end{tabular}
\end{ruledtabular}
\end{table*}

The first eight Mellin moments of pseudo-scalar and vector mesons are listed in Table\,\ref{tab:PS_all}, \ref{tab:PX_all}, \ref{tab:CZ_all}.
 
\begin{acknowledgments}

 We would like to thank Zhen-Ni Xu, Zhao-Qian Yao, Kh\'epani Raya, Jorge Segovia, Jos\'e Rodríguez-Quintero and Craig D. Roberts for useful discussions. This work has been partially funded by Ministerio Espa\~nol de Ciencia e Innovaci\'on under grant Nos. PID2019-107844GB-C22 and PID2022-140440NB-C22; Junta de Andaluc\'ia under contract Nos. Operativo FEDER Andaluc\'ia 2014-2020 UHU-1264517, P18-FR-5057 and also PAIDI FQM-370. The authors acknowledge, too, the use of the computer facilities of C3UPO at the Universidad Pablo de Olavide, de Sevilla.
\end{acknowledgments}

\bibliography{ref.bib}

\end{document}